\documentclass[10pt,journal,cspaper,compsoc]{IEEEtran}

\usepackage{mathptmx}
\usepackage{graphicx}
\usepackage{times}
\usepackage[lined,algonl,boxed]{algorithm2e}
\usepackage{amssymb,amsmath}
\usepackage{multirow}
\usepackage{subfigure}
\usepackage{color}
\usepackage{paralist}
\usepackage{tikz}
\usepackage{multirow}
\usepackage{color}
\usepackage{ulem}
\usepackage{setspace}
\usepackage{microtype,hyphenat,balance}
\usepackage{stfloats}
\usepackage{array}
\usepackage{url}
\usepackage{bm}
\usepackage[singlelinecheck=false]{caption}
\usepackage{soul}
\newcommand{\xiting}[1]{\textcolor{black}{#1}}

\newcommand{\kg}[1]{\textcolor{black}{#1}}
\newcommand{\dc}[1]{\textcolor{black}{#1}}
\newcommand{\docpr}[1]{\textcolor{black}{#1}}
\newcommand{\docat}[1]{\textcolor{black}{#1}}
\newcommand{\tvcgminor}[1]{\textcolor{black}{#1}}
\newcommand{\nop}[1]{}

\title{Online Visual Analytics of Text Streams}

\author{ Shixia Liu, Jialun Yin, Xiting Wang, Weiwei Cui, Kelei Cao, Jian Pei
\IEEEcompsocitemizethanks{
\IEEEcompsocthanksitem S. Liu is with School of Software, Tsinghua University.\protect\\
E-mail: shixia@tsinghua.edu.cn.
\IEEEcompsocthanksitem Jialun Yin, Xiting Wang, and Kelei Cao are with Tsinghua University.\protect\\
E-mail: \{yinjl14, wang-xt11, ckl13\}@mails.tsinghua.edu.cn.

\IEEEcompsocthanksitem Weiwei Cui is Microsoft Research.
E-mail: weiwei.cui@microsoft.com.

\IEEEcompsocthanksitem Jian Pei is with Simon Fraser University, Burnaby, BC Canada.\protect\\
E-mail: jpei@cs.sfu.ca.

}

\thanks{}}

\IEEEcompsoctitleabstractindextext{%
\begin{abstract}
We present an online visual analytics approach to helping users explore and understand hierarchical topic evolution in high-volume text streams.
The key idea behind this approach is to identify representative topics \docpr{in incoming} documents and align them with the existing representative topics that they immediately follow (in time).
To this end, we learn a set of streaming tree cuts from topic trees based on user-selected focus nodes.
A dynamic Bayesian network model \docpr{has been} developed to derive the tree cuts in the \dc{incoming} topic trees \kg{to balance} the fitness of each tree cut and the smoothness between adjacent tree cuts.
By connecting the corresponding topics at different times, we \docpr{are able to} provide an overview of the evolving hierarchical topics.
A sedimentation-based visualization \dc{has been} designed to enable the interactive analysis of streaming text data from global patterns to local details.
We \docpr{evaluated} our method on real-world datasets and the results are generally favorable.

\end{abstract}

} 

\begin{document}

\maketitle
{
\fontsize{8}{8} 

\section{Introduction}

Surveying and exploring text streams that have many hierarchical and evolving topics are important aspects of many big data applications~\cite{Chakrabarti2006,Wang2013}.
For example, \docpr{the use of} such evolving hierarchical topics \docpr{allows for the detection} and \docpr{tracking of} new and emerging events (e.g., Ebola outbreak) in a huge volume of streaming news articles and microblog posts.
Exciting progress, such as learning topics from text streams, has been made in mining text streams~\cite{ Wang2013}.
However, one essential problem remains: how can we effectively present interesting topics and track their evolution over time in a comprehensible and manageable manner?
This task is a key to \docpr{connecting} big data with people.\looseness=-1

Let us consider an example to understand this challenge.
Suppose an analyst reads an article entitled \docpr{``}Third U.S. Aid Worker Infected with Ebola Arrives in Nebraska.\docpr{''}
\docat{\docpr{The analyst} is interested in the topic ``Ebola-infected aid workers'' and wants to analyze the relevant discussions in the \docpr{subsequent weekly} news \docpr{articles}.}
In addition, s/he is interested in how this topic is related \docpr{to other} topics in the news stream as time progresses, especially the newly generated topics.
Such analysis helps the analyst understand the relationship between the severity of Ebola and the intensity of public opinion.
Based on this relationship, s/he can \docpr{make suggestions to the government}.

A text stream, such as the aforementioned Ebola dataset, often contains hundreds or even thousands of topics that can be naturally organized in a tree, \docat{\docpr{known as} a topic tree~\cite{Blundell2010,Wang2013,Zhang2009}}.
A topic tree may change as new documents arrive.
We can mine a sequence of coherent topic trees to represent major topics in the text stream and their evolution over time~\cite{Wang2013}.
However, the question of whether such a sequence of topic trees is effective enough to analyze and understand \docat{a text stream \docpr{remains,}
\docpr{in} particular,} whether these topic trees can illustrate the accumulation and aggregation of the new documents into the existing topics.


To address this problem, we have developed a visual analytics system, \emph{\normalsize TopicStream}, to help users explore and understand hierarchical topic evolution in a text stream.
Specifically, we incrementally extract a new tree cut from the incoming topic tree, based on a dynamic Bayesian network (DBN) model.
We model the topics that a user is interested in as proper tree cuts in a sequence of topic trees similar to~\cite{cui2014}.
A tree cut is a set of tree nodes describing the layer of topics that a user is interested in.
In \emph{\normalsize TopicStream}, we employ the DBN model to derive the tree cut \docpr{from} an incoming topic tree.
A time-based visualization is then developed to present the hierarchical clustering results and their alignment over time.
In particular, \docpr{we have adopted} a customized sedimentation \docpr{metaphor to} visually illustrate how incoming text documents are aggregated over time into the existing document archive, including document entrance into the scene from an entrance point, suspension while approaching to the topic, accumulation and decay on the topic, as well as aggradation with the topic over time~\cite{Wang2013}.

We make the following technical contributions in this work:\looseness=-1
\begin{compactitem}
\item \textbf{\normalsize A streaming tree cut algorithm} is proposed to extract an optimal tree cut for an incoming topic tree based on user interests. This algorithm produces a sequence of representative topic sets for different topic trees, which smoothly evolve over time.
\item \textbf{\normalsize A sedimentation-based metaphor} is integrated into the river flow metaphor to visually illustrate how new documents are aggregated into old documents. It helps analysts immediately track and understand incoming topics and connect those topics with existing ones.
\item \textbf{\normalsize A visual analytics system} is built to integrate evolutionary hierarchical clustering ~\cite{Wang2013} and the streaming tree cut techniques into an interactive visualization. The unique feature of this system is its ability to provide a coherent view of evolving topics in text streams.
\end{compactitem}


\section{Related Work}\label{sec:related-work}

\begin{figure*}[t]
  \centering

  \includegraphics[width=\linewidth]{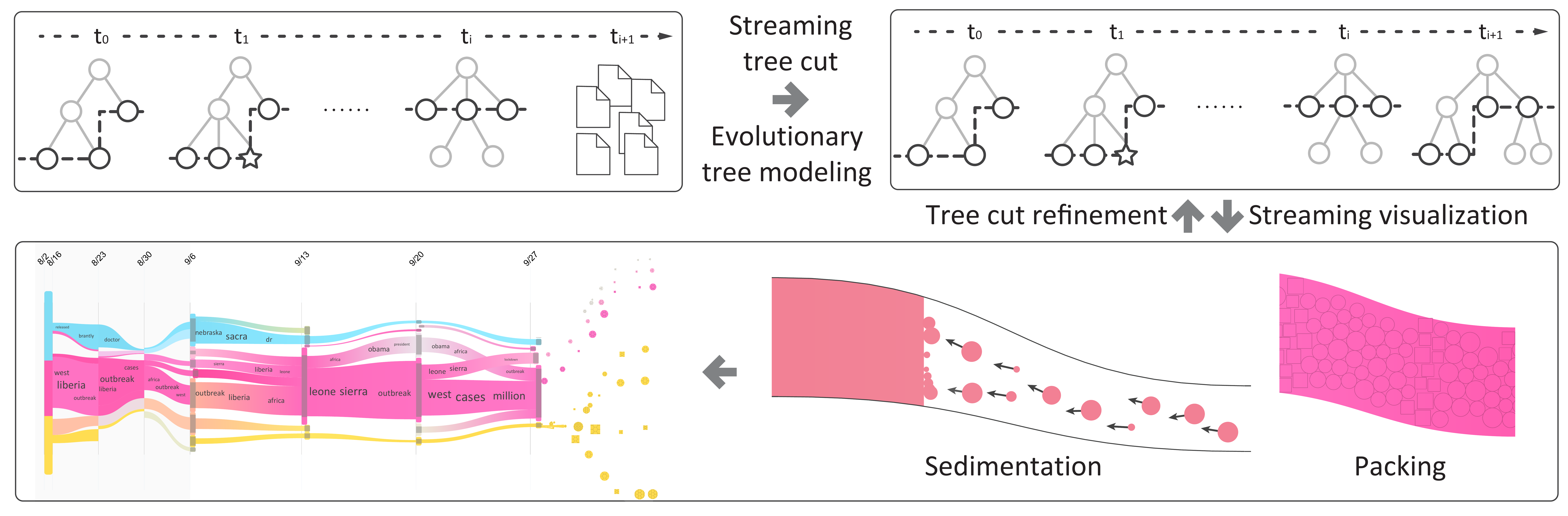}
  \vspace{-5mm}
  \caption{\emph{\normalsize TopicStream} system consists of two modules: streaming tree cut and streaming visualization.}
  \label{fig:overview}
  \vspace{-2mm}
\end{figure*}

\subsection{Evolutionary Topic Analysis}
Various generative-probabilistic-model-based machine learning algorithms, such as dynamic Latent Dirichlet Allocation (LDA)~\cite{blei2006} and hierarchical Dirichlet processes~\cite{Ahmed2008,Ahmed2010,Xu2008a,Zhang2010}, have been proposed to extract evolving topics from a text stream.
MemeTracker~\cite{Leskovec2009} was developed to effectively identify phrase-based topics from millions of news articles.
In many applications, evolving topics may be related to or correspond to one another over time.
The most intuitive relationships are \emph{\normalsize topic correlation}~\cite{wang2007} and \emph{\normalsize common topics}~\cite{wang2009}.
Recent efforts have focused on the analysis of topic evolution patterns in text data, including topic birth, death, splitting, and merging~\cite{Gao2011}.
Although the aforementioned methods help users understand a text corpus, none of them focus on mining and understanding streaming hierarchical topics.

Some efforts have also been exerted recently to mine hierarchical topics and their evolving patterns in temporal datasets.
The evolutionary hierarchical clustering algorithm~\cite{Chakrabarti2006} generates a sequence of hierarchical clusters.
The major feature of this algorithm is that clustering properly fits current data at any time (fitness). Furthermore, clustering does not shift dramatically from one time step to the next when content is similar (smoothness).
However, this algorithm can only generate evolving binary trees.
To tackle this issue, Wang et al.~\cite{Wang2013}
formulated the multi-branch tree construction problem as  a Bayesian online filtering process.
Unlike the method proposed in~\cite{Wang2013}, our method addresses the problem of better understanding and analyzing a sequence of evolutionary multi-branch topic trees.
We first learn a set of evolutionary tree cuts from the topic trees based on the user-selected focus nodes.
Then we design a sedimentation-based interactive visualization to reveal hierarchical topic evolution from multiple perspectives.


\subsection{Visual Topic and Event Evolution}

The visual analysis of evolving topics in text corpora has been widely studied in recent years~\cite{cui2011,Havre2002,Liu2014survey,Sun2013survey}.
Many methods utilize a river metaphor (a stacked graph) to convey evolving topics over time.
For example, ThemeRiver~\cite{Havre2002} visually depicts how keyword strengths change over time in a text corpus through a river metaphor.
A layer represents a topic in this metaphor.
The varying width of a layer represents \dc{strength} change over time.
TIARA~\cite{Liu2009,Liu2012}
employs an enhanced stacked graph to illustrate how topics evolve over time.
ParallelTopics~\cite{Dou2011} utilizes ThemeRiver to illustrate topic evolution over time and parallel coordinate plots to reveal the probabilistic distribution of a document on different topics.
TextFlow~\cite{cui2011} was developed to help analysts visually analyze topic merging and splitting relationships and track their evolution over time.
A visual analysis system was designed by Xu et al.~\cite{Xu2013} to allow analysts to interactively explore and understand the dynamic competition relationships among topics.
Recently, Sun et al.~\cite{GSun2014a} extended this work to study both the cooperation and competition relationships among topics.\looseness=-1

Several visualization techniques have been proposed recently to help users analyze temporal events and their evolving patterns~\cite{Dou2012,krstajic2013,Luo2012}.
EventRiver~\cite{Luo2012} assumes that clusters of news articles with similar content are adjacent in time and can be mapped to events.
Thus, this method automatically detects important events and visually presents their impact over time.
LifeFlow~\cite{Wongsuphasawat2011} and Outflow~\cite{Wongsuphasawat2012} help users explore temporal event sequences.


The aforementioned approaches focus on the visual exploration of evolving topics/events with flat structures.
By contrast, our method attempts to support the visual analysis of evolving hierarchical topics over time.

HierarchicalTopics~\cite{Dou2013} hierarchically organizes topics using the BRT model~\cite{Blundell2010,Liu2012a}, which can then represent a large number of topics without being cluttered.
However, this method utilizes one static tree to organize all topics and cannot illustrate splitting/merging relationships among topics.\looseness=-1

To solve this problem, Cui et al.~\cite{cui2014} developed \emph{\normalsize RoseRiver} to progressively explore and analyze the complex evolution patterns of hierarchical topics.
This system introduces the concept of evolutionary tree cuts to help better understand large document collection with time-stamps.
However, it fails to provide a mechanism to analyze streaming data because a global tree cut algorithm is employed.
In addition, the authors used the DOI-based heuristic rule to derive the key tree cut, which may not be the optimal solution.

Unlike the preceding method, we employ a-posterior-probability-based method to estimate the fitness of the tree cut.
We then formulate the derivation of the tree cut in incoming data as a DBN.
The quantitative evaluation in Sec.~\ref{sec:quantitativeevaluation} shows that the posterior-probability-based method performs better than the DOI-based method in~\cite{cui2014} to fit the focus nodes and topic trees.
The performance of the DBN-based streaming tree cut algorithm is comparable with that of the global tree cut algorithm proposed in~\cite{cui2014}.
These observations demonstrate the effectiveness of the proposed mining algorithms at handling incoming data in text streams.
An improved visual sedimentation metaphor~\cite{Huron2013visual} has been adopted to visually illustrate how incoming text streams aggregate into existing topics.



\section{System Overview}

\emph{\normalsize TopicStream} is designed to track and understand the dynamic characteristics of text streams.
\kg{It} consists of two major modules: streaming tree cut and streaming visualization (Fig.~\ref{fig:overview}).\looseness=-1

The input of the streaming tree cut is a set of topic trees with tree cuts \kg{and} a set of \kg{incoming} documents.
In \emph{\normalsize TopicStream}, the topic trees are derived by the evolutionary tree clustering method developed by Wang et al.~\cite{Wang2013}.
The basic idea of this method is to balance the fitness of a tree and the smoothness between trees by a Bayesian online filtering process.
We derive the tree cuts based on the user-selected focus node(s).
This module \kg{initially} extracts a topic tree from the newly arrived documents \kg{using} the evolutionary tree clustering model~\cite{Wang2013}.
\kg{A} tree cut is \kg{then} derived from the new topic tree \kg{through} the developed streaming tree cut algorithm.\looseness=-1

The streaming tree cuts are \kg{then} fed into the visualization module.
We employ the visual sedimentation metaphor to reveal the merging process of newly arrived documents \kg{with} the dominant center of visualization.
The circle packing algorithm is also developed to illustrate the relationships of document clusters within each topic stripe, including their similarity and temporal relationships~\cite{Wang2006visualization,ZhaoTVCG2014}.\looseness=-1

\section{Streaming Tree Cut}\label{sec:tree-cut}
\kg{This section explains} how \kg{a} new tree cut \kg{is} incrementally \kg{derived} as new text data arrives.


\subsection{Problem Formulation}

\kg{We} use \docpr{a} tree cut to represent a topic tree based on user interests, which is similar to~\cite{cui2014}.
A \textbf{\normalsize tree cut} is a set of nodes \dc{in which} every path from the root of the tree to a leaf contains exactly one node from the cut.
Thus, each cut can be used as a set of representative topic nodes.
\kg{That is}, a tree cut represents a level of topic granularity of a user's \docpr{interests}.
Fig.~\ref{fig:treecutexample} represents an example of a tree cut.
We refer to each node on the tree cut as a ``\textbf{\normalsize cut node}.''



\begin{figure}[t]
  \centering
  \includegraphics[width=2.5in]{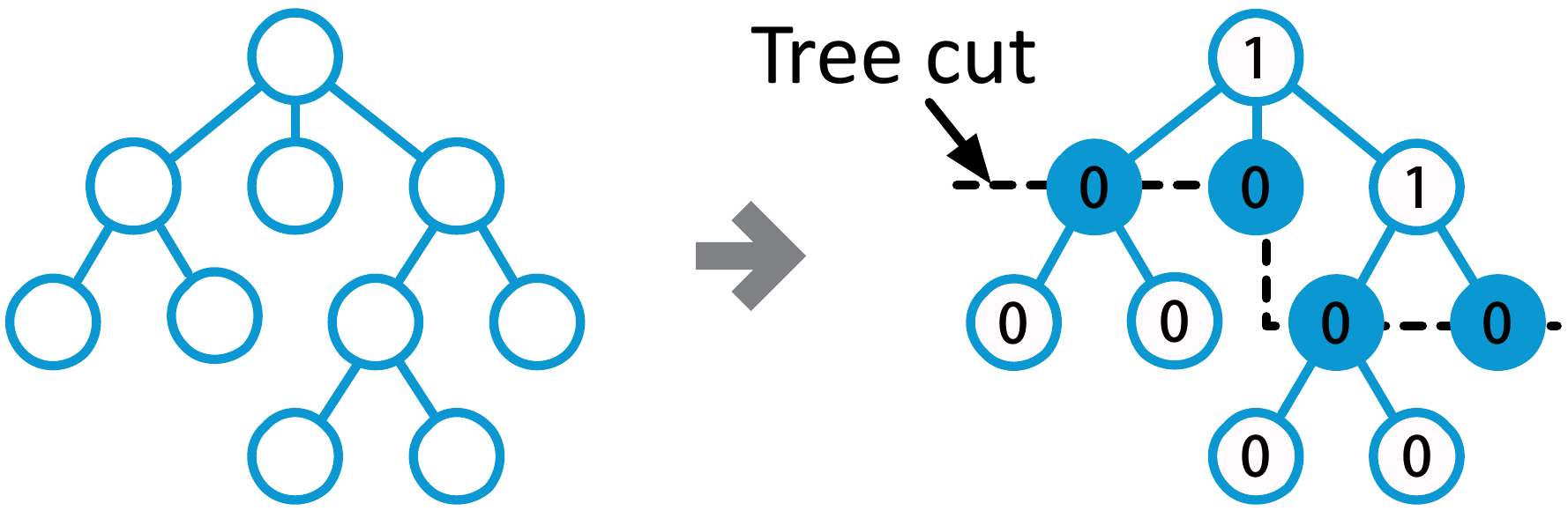}
  \caption{
Tree cut example: the cut is denoted by the dotted line. Every node above the dotted line is labeled 1, while the \docpr{others} are labeled 0.
}
  \label{fig:treecutexample}
  \vspace{-5mm}
\end{figure}

The basic principle \kg{to determine} a set of optimal tree cuts is that each tree cut in the sequence should be similar to the one at the previous time step if the tree structures are similar (smoothness).
The tree cut must also adequately represent user interests and the topic tree at that time step (fitness).
The global tree cut algorithm developed in~\cite{cui2014} computes all tree cuts \kg{simultaneously} based on the focused nodes.
\kg{Two} problems arise when applying \kg{the aforementioned algorithm} to the text stream.
First, it is very time consuming \docpr{to} compute all the tree \docpr{cuts each time} new text data arrives.
Second, if the tree cuts are \docpr{recomputed along} with \dc{the} new data, \kg{then} the existing tree cuts \kg{are} changed to \kg{a certain} extent, which \kg{makes maintaining the} mental map \dc{of analysts difficult}~\cite{Woods1984}.

\begin{figure}[t]
  \centering
  \includegraphics[width=2.1in]{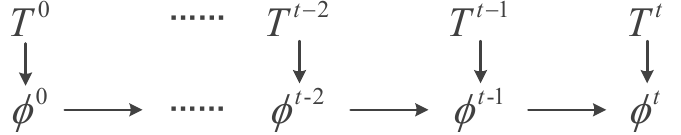}
  \caption{
  Dynamic Bayesian network for deriving streaming tree cuts. Here $T^t$ and ${\phi}^t$ are the topic tree and tree cut at \emph{t}.
  }\label{fig:hmm}
  \vspace {-5mm}
\end{figure}

\begin{table}[b]
\small
    \vspace{-5mm}

    \centering
    \scalebox{0.8}{

    \begin{tabular}{|c|c|}
    \hline
    \textbf{Notation} & \textbf{Definition} \\
    \hline
   	$T^t$ & The topic tree at time $t$ \\
   \hline
	$\phi^t$ & The tree cut at time $t$ \\
	\hline
	$m$ & Number of focus nodes selected by the user \\
	\hline
	$T_{fi}$ &  The $i$th focus node \\
	\hline
	$\mathcal{D}_{fi}$ &  The document set of the $i$th focus node\\
	\hline	
	$ p({\phi}^t|{\phi}^{t-1},T^t)$ & The conditional distribution of $\phi^t$ according to DBN \\
	\hline
	$p({\phi}^t|T^t)$ & \tvcgminor{Fitness of tree cut ${\phi}^t$ to $T^t$} \\
	\hline
	$p({\phi}^t|{\phi}^{t-1})$ & Smoothness between adjacent tree cuts $\phi^t$ and $\phi^{t-1}$ \\
   \hline	
   	\tvcgminor{$p({\phi}^t|\mathcal{D}_{f0}, ..., \mathcal{D}_{fm})$} & The posterior probability of a tree cut ${\phi}^t$ \\
	\hline   	
	$E_1(T^t)$ & The similarity energy of the topic tree $T^t$ \\
   	\hline	
	$T_r$, $T_s$ &  A topic node (an internal node) in the topic tree \\
	\hline	   	
   	$\mathbf{S}(T_r,T_s)$ & The cosine similarity between topic nodes $T_r$ and $T_s$ \\
   	\hline	
	$l_s$ &  The label (0 or 1, see Fig.~\ref{fig:treecutexample}) of topic node $T_s$  \\
	\hline	
   	$E_2(\phi^{t}, \phi^{t-1})$ & The smoothness energy between tree cuts $\phi^t$ and $\phi^{t-1}$ \\
   	\hline	   	
	$\mathcal{D}_s$ &  The document set of topic node $T_s$\\
	\hline	
    $f_{DCM}(\mathcal{D})$  & The marginal distribution of document set $\mathcal{D}$ \\	
	\hline
	$\mathbf{WS}(T_c)$ &  Window size for $T_c$ in mean-shift clustering \\
	\hline
    \end{tabular}
    }
       \caption{
    \tvcgminor{Frequently used notations in the model.} 
    }
     \label{table:notations}
\end{table}

To solve \kg{the aforementioned} problems, we \docpr{have adopted} a dynamic Bayesian network (DBN) model to infer the tree cut for the incoming text data organized by a topic tree.
Previous studies have shown that adopting overlapping successive views to support continuity across data sets is a frequently adopted principle \kg{to process} temporal data~\cite{Chakrabarti2006,Woods1984}.
In our case, the new tree cut ${\phi}^t$ is relevant to temporally \kg{adjacent} tree cuts as well as $T^t$ (Fig.~\ref{fig:hmm}).
\kg{In particular}, topic mapping between adjacent trees is utilized as a constraint to \dc{smooth} tree cut transitions over time.


\subsection{Model}



\begin{figure*}[t]
\centering
\includegraphics[width=0.9\textwidth]{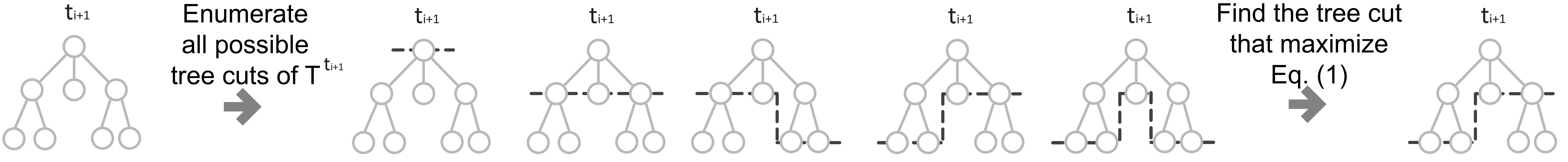}
\caption{\tvcgminor{Streaming tree cut algorithm: given the incoming topic tree $T^{t_{i+1}}$, we enumerate all possible tree cuts of $T^{t_{i+1}}$ and then pick the tree cut that maximizes Eq.~(\ref{eq:hmm1}).}}
\vspace{-5mm}
\label{fig:treecutalgorithm}
\end{figure*}

Assume \kg{that} we already have a sequence of topic trees and the corresponding tree cuts.
\kg{The} problem of deriving \kg{a} new tree cut in a text stream can \kg{then} be regarded as a labeling problem.
\kg{The} topic nodes above the tree cut are labeled 1\kg{, whereas} the rest (including the cut nodes) are labeled 0 (Fig.~\ref{fig:treecutexample}).
We first introduce some frequently used notations in Table~\ref{table:notations}, which are useful for subsequent discussions.

Given \emph{\normalsize m} focus nodes \{${T_{fi}}$\} with document sets \{$\mathcal{D}_{fi}$\}, we infer the tree cut ${\phi}^t$ in the \dc{incoming} topic tree $T^t$.
Fig.~\ref{fig:hmm} \kg{shows that} $T^t$ is an observation \kg{variable} and ${\phi}^t$ is a hidden variable.
The relationship between ${\phi}^t$ and \docpr{${\phi}^{t-1}$, as well as $T^t$, can} be modeled by DBN.
Accordingly, the conditional distribution of ${\phi}^t$ is $p({\phi}^t|{\phi}^{t-1},T^t)$.
Since ${\phi}^t$ is relevant to $\mathcal{D}_{f0},\mathcal{D}_{f1}, ..., \mathcal{D}_{fm}$ at each time $t$, we formulate the inference of the new tree cut as:

\vspace{-5mm}
\begin{equation}
max\ p({\phi}^t,{\phi}^{t-1}, ...,{\phi}^0|\mathcal{D}_{f0},\mathcal{D}_{f1}, ..., \mathcal{D}_{fm})\cdot p({\phi}^t|{\phi}^{t-1},T^t).
\label{eq:hmm1}
\end{equation}


\tvcgminor{As shown in Fig.~\ref{fig:treecutalgorithm}, the goal is to find the tree cut that maximizes Eq.~(\ref{eq:hmm1}).
}

Since ${\phi}^t, {\phi}^{t-1}, ..., {\phi}^0$ are conditionally independent given $\mathcal{D}_{f0},\mathcal{D}_{f1}, ..., \mathcal{D}_{fm}$,
the first term is computed by $\prod\limits_{\tau=0}^{t}p({\phi}^\tau|\mathcal{D}_{f0},\mathcal{D}_{f1}, ..., \mathcal{D}_{fm})$.
According to the graphical model of DBN (Fig.~\ref{fig:hmm}), the second term is proportional to $p({\phi}^{t}|T^{t})p({\phi}^{t}|{\phi}^{t-1})$.
\xiting{Because} ${\phi}^{t-1},\ {\phi}^{t-2},...,\ {\phi}^{0}$ are known, Eq.~(\ref{eq:hmm1}) can be simplified as:

\vspace{-4mm}
\begin{equation}
max\ p({\phi}^t|T^t)p({\phi}^{t}|{\phi}^{t-1})p({\phi}^{t}|\mathcal{D}_{f0},\mathcal{D}_{f1}, ..., \mathcal{D}_{fm}).
\label{eq:hmm2}
\end{equation}

$p({\phi}^t|T^t)$ denotes how well the tree cut ${\phi}^t$ represents $T^t$, which is measured by the similarity energy $\mathbf{E}_1$ in the form of $p({\phi}^t|T^t) = e^{-E_1(T^t)}$.
$\mathbf{E}_1$ measures the content similarity of each topic $T_r$ in $T^t$ \docpr{for} the two topic sets, which are topic \docpr{node} sets labeled 0 and \docpr{1, respectively}.

\vspace{-3mm}
\begin{equation}
\small
E_1(T^t) = \sum_{T_r\in{\mathcal{N}^t}}\min_{T_s\in{\mathcal{N}^t},l_s=l_r}(-\log(\mathbf{S}(T_r,T_s))),
\end{equation}
where $l_r$ is the label (1 or 0) of topic node $T_r$, $\mathcal{N}^t$ is the set which contains all tree nodes in $T^t$.
For a topic $T_r$, \kg{the} cosine similarity $\mathbf{S}(T_r,T_s)$ is used to compute the similarity value between $T_r$ and $T_s$ with the same label.

$p({\phi}^t|{\phi}^{t-1})$ measures the smoothness cost between two adjacent tree cuts using the smoothness energy $\mathbf{E_2}$, which is defined as $p({\phi}^t|{\phi}^{t-1})= e^{-E_2(\phi^t,\phi^{t-1})}$. $\mathbf{E_2}$ measures the mapping similarity between $T^t$ and $T^{t-1}$:
\vspace{-3mm}
\begin{equation}
E_2(\phi^t,\phi^{t-1}) = \sum\limits_{T_r\in{\mathcal{N}^t},T_s\in{\mathcal{N}^{t-1}}}|l_r-l_s|\varphi(l_r,l_s),
\label{eq:hmm5}
\end{equation}
where $\varphi(l_r,l_s)$ denotes the mapping weight computed by the evolutionary tree clustering model.

%

$p({\phi}^t|\mathcal{D}_{f0},\mathcal{D}_{f1}, ..., \mathcal{D}_{fm})$ is defined as a posterior probability of a tree cut ${\phi}^t$\kg{. Thus,}

\vspace{-5mm}
\begin{equation}
\label{eq:seed}
p({\phi}^t|\mathcal{D}_{f0}, \mathcal{D}_{f1}, ..., \mathcal{D}_{fm}) \propto p(\mathcal{D}_{f1}, \mathcal{D}_{f2}, ..., \mathcal{D}_{fm}|{\phi}^t) p({\phi}^t),
\end{equation}
where $p(\mathcal{D}_{f0}, \mathcal{D}_{f1}, ..., \mathcal{D}_{fm}|{\phi}^t)$ is the likelihood of the tree cut.
$p({\phi}^t)$ indicates the prior preference of the node number for the tree cut.
The tree cut that results in maximum posterior probability is the optimal tree cut.

$p({\phi})$ is defined as $e^{-\lambda|\mathcal{C}_{{\phi}}|}$, where $\mathcal{C}_{{\phi}}$ is the set of topics in ${\phi}$, $|\mathcal{C}_{\phi}|$ is the node number in the tree cut, and $\lambda$ is the parameter that balances the
likelihood and expected node number.\looseness=-1

We then illustrate how the likelihood of a tree cut \kg{can be calculated}.
We adopt a prediction model to estimate the likelihood of \kg{each} possible tree cut.
For simplicity\dc{'s sake}, we begin with one focus node. 
Given a focus node $T_f$ and its corresponding document set $\mathcal{D}_f$, the predictive probability of a tree cut $\phi$ is defined as: 
\begin{equation}
p(\mathcal{D}_f|\phi) = \sum_{T_s\in \mathcal{C}_{\phi}} \omega_s p(\mathcal{D}_f|\mathcal{D}_s),
\end{equation}
where $\mathcal{C}_{\phi}$ is the set of topics in $\phi$ and $\omega_s$ is the prior probability that all the documents in $\mathcal{D}_f$ belong to $\mathcal{D}_s$.
To calculate $\omega_s$, we assume that the probability of a set of documents belonging to a specific topic is proportional to the number of documents in that topic~\cite{Blundell2011}.
Accordingly, we obtain $\omega_s = {|\mathcal{D}_s|}/{|\mathcal{D}_a|}$.
$\mathcal{D}_a$ includes all documents in \kg{a} tree.
$p(\mathcal{D}_f|\mathcal{D}_s)$ is the predictive distribution of the corresponding \xiting{topic model.}
\begin{equation}
\small
p(\mathcal{D}_f|\mathcal{D}_s) = {f(\mathcal{D}_f \cup \mathcal{D}_s)}/{f(\mathcal{D}_s)},
\end{equation}
where $f(\mathcal{D})$ is the marginal probability of data $\mathcal{D}$.

\kg{The} Dirichlet compound multinomial (DCM) distribution is derived \kg{from} multinomial and Dirichlet conjugate distributions~\cite{Liu2012a}.
\dc{Because} \kg{it} relies on hierarchical Bayesian modeling techniques, \kg{DCM} is a more appropriate generative model than the traditional multinomial distribution for text documents.
Thus, we utilize the DCM distribution to represent the marginal distribution $f(\mathcal{D})$ \kg{as follows:}
\begin{equation}
\label{eq:dcm}
f_{DCM}(\mathcal{D}) =  \prod_i^n \frac{(\sum_j^{|\mathcal{V}|}z_i^{(j)})!}{\prod_j^{|\mathcal{V}|} z_i^{(j)}!} \cdot \frac{\Delta({\bm \alpha}+\sum_i\bold{z}_i)}{\Delta({\bm \alpha})},
\end{equation}
where \xiting{$|\mathcal{V}|$ is the vocabulary size, $\bold{z}_i\in \mathcal{R}^{|\mathcal{V}|}$ is the word vector of the $i$th document, and $z_i^{(j)}$ is the frequency of the $j$th term.} 
${\bm \alpha} = (\alpha^{(1)},\alpha^{(2)},\ldots,\alpha^{(|\mathcal{V}|)})^T \in \mathcal{R}^{|\mathcal{V}|}$ is the parameter that controls the Dirichlet distribution, which is the prior of the multinomial distribution of each topic.
$\Delta({\bm \alpha})$ is the Dirichlet delta function defined by $\Delta({\bm \alpha}) =\Gamma(\sum_{j=1}^{|\mathcal{V}|} \alpha^{(j)})/\prod_{j=1}^{|\mathcal{V}|} \Gamma(\alpha^{(j)})$.
\kg{The gamma} function has the property $\Gamma(\alpha+1) = \alpha\Gamma(\alpha)$.

We then extend the likelihood formulation to any number of focus nodes.
When several focus nodes are selected, the predictive probability of a tree cut is \kg{as follows:}
\begin{equation}
\small
p(\mathcal{D}_{f1}, \mathcal{D}_{f2}, ..., \mathcal{D}_{fm}|\phi) = \prod_{i\in [1, m]} p(\mathcal{D}_{fi}|\phi).
\end{equation}
Directly maximizing the \kg{aforementioned} predictive probability is intractable; thus, we adopt the tree pruning procedure \kg{presented in}~\cite{He2003} for optimal tree cut selection.


\subsection{Postprocessing}
\kg{A} set of representative topic nodes is selected to represent the topic tree at each time step.
\kg{using the evolving tree cut algorithm}.
However, two issues remain.
First, the resulting tree cuts \kg{may not} ideally reflect \kg{user interests} because a topic node \kg{can} have any number of children.
For example, a topic node that is highly related to the focus node \kg{can} have many \kg{less-related} siblings.
Considering that a tree cut cannot \kg{simultaneously} include \kg{a} highly related node and its parent, all \kg{of} its siblings have to be included in the tree cut as well.
This condition \kg{results} in showing less-related content with unnecessary details.
Second, the number of representative topics at \kg{several} time steps is still too large to be displayed in the limited screen area \kg{in many cases}.\looseness=-1

To address these issues, we propose a postprocessing approach to further reduce the topic number.
\kg{This approach} (1) \kg{encourages} the merging of related siblings with similar \dc{content that is less} related to the focus nodes, (2) \kg{discourages} the merging of topics that \kg{are} highly related to the focus nodes, and (3) \kg{maintains} smoothness between adjacent topic sets over time.\looseness=-1

To meet the aforementioned requirements, a clustering method is needed.
Mean-shift clustering~\cite{comaniciu2002mean}, which automatically determines the cluster number, can \kg{be easily} adapted to fulfill all the requirements.
The first requirement can be satisfied by any clustering method. Thus, we focus on how to fulfill the \kg{remaining requirements}.\looseness=-1

To meet the second requirement, an adaptive window size $\mathbf{WS}$ is defined for different clustering centers $T_c$.
\begin{equation}
\small
	\mathbf{WS}(T_c)=
		\begin{cases}
			0 & \text{if } \mathbf{S}(\mathcal{D}_c, \mathcal{D}_f)\ge \gamma,\\
			{(\gamma-\mathbf{S}(\mathcal{D}_c, \mathcal{D}_f))w_{max}/\gamma} & \text{otherwise}.
		\end{cases}
\end{equation}
where $\gamma$ is the similarity threshold, $w_{max}$ is the maximum window size, and $\mathbf{S}(\mathcal{D}_c, \mathcal{D}_f)$ is the cosine similarity.

To meet the third requirement, all the tree cuts are generated in temporal order.
Smoothness between adjacent topic sets is preserved by treating the previous clustering centers as the initial centers of the current cut node clustering.\looseness=-1



\section{Visualization}
\label{sec:vis}

\subsection{Design Rationale}
\label{sec:rationale}
We designed the \emph{\normalsize TopicStream} visualization iteratively with three domain experts, including one professor in media and communication (P1), one professor \dc{who} majored in public opinion analysis \dc{in} healthcare (P2), and one researcher who \kg{operates} a visualization \kg{start-up} (S3).
These experts are not co-authors of this paper.
We discussed with the experts about the analysis process and need in their work.
\kg{In general,} they \dc{desire} a system that provides a coherent view of the evolving topics in text streams and compares \kg{incoming} content with \kg{previous} \dc{content}.
We derived the following design guidelines \kg{based on their feedback and previous \dc{research}.}

\noindent \textbf{\normalsize R1 - Providing an overview of a text stream}.
The experts requested \kg{a summary of old, recent, and incoming documents} in the text stream.
With \dc{such a summary}, they can easily form a full picture of the text stream, including its major topics \kg{and} their evolutionary patterns over time.
In addition, \kg{a} summary was also requested to provide historical and contextual information for \kg{incoming} documents.
This is consistent with the design rationale of fisheye view~\cite{Furnas1986}.
Expert S3 commented that, ``\dc{a} smooth transition between new data and old data is very helpful for me \dc{to find connections}.''\looseness=-1

\noindent \textbf{\normalsize R2 - Revealing how \kg{incoming} documents merge with existing ones}.
\kg{Previous} research \dc{into} visual sedimentation~\cite{Huron2013visual} has shown that \dc{a} smooth transition between the focus (new data) and the context (old data) \dc{helps} users understand \kg{a} text stream.
\kg{The} experts also confirmed that understanding how \kg{incoming} documents \kg{merge} with historical documents is useful \kg{in} their analysis.
For example, P1 said \kg{that}, ``\kg{Examining} the speed, volume, and sequential order \kg{of incoming data} is very useful \kg{to study} \dc{agenda setting} in my field.''\looseness=-1


\noindent \textbf{\normalsize R3 - Comparing document content at different times}.
\kg{Experts frequently} compare the content of new documents with \kg{those} of old ones in their daily analysis.
For example, expert P2 commented \kg{that}, ``\kg{In} a multi-source text stream, one source may \kg{follow} another to publish documents on a specific topic.
I am interested in comparing this follower-followee relationships in the new time slot with that of other time stpes, \kg{to obtain} a clear understanding \kg{of} who follows whom in a topic of interest.''
\kg{Thus} the system \kg{should} facilitate the visual comparison of documents at different times.\looseness=-1

\begin{figure}[b]
	\centering
\vspace{-4mm}
	\includegraphics[width=\columnwidth]{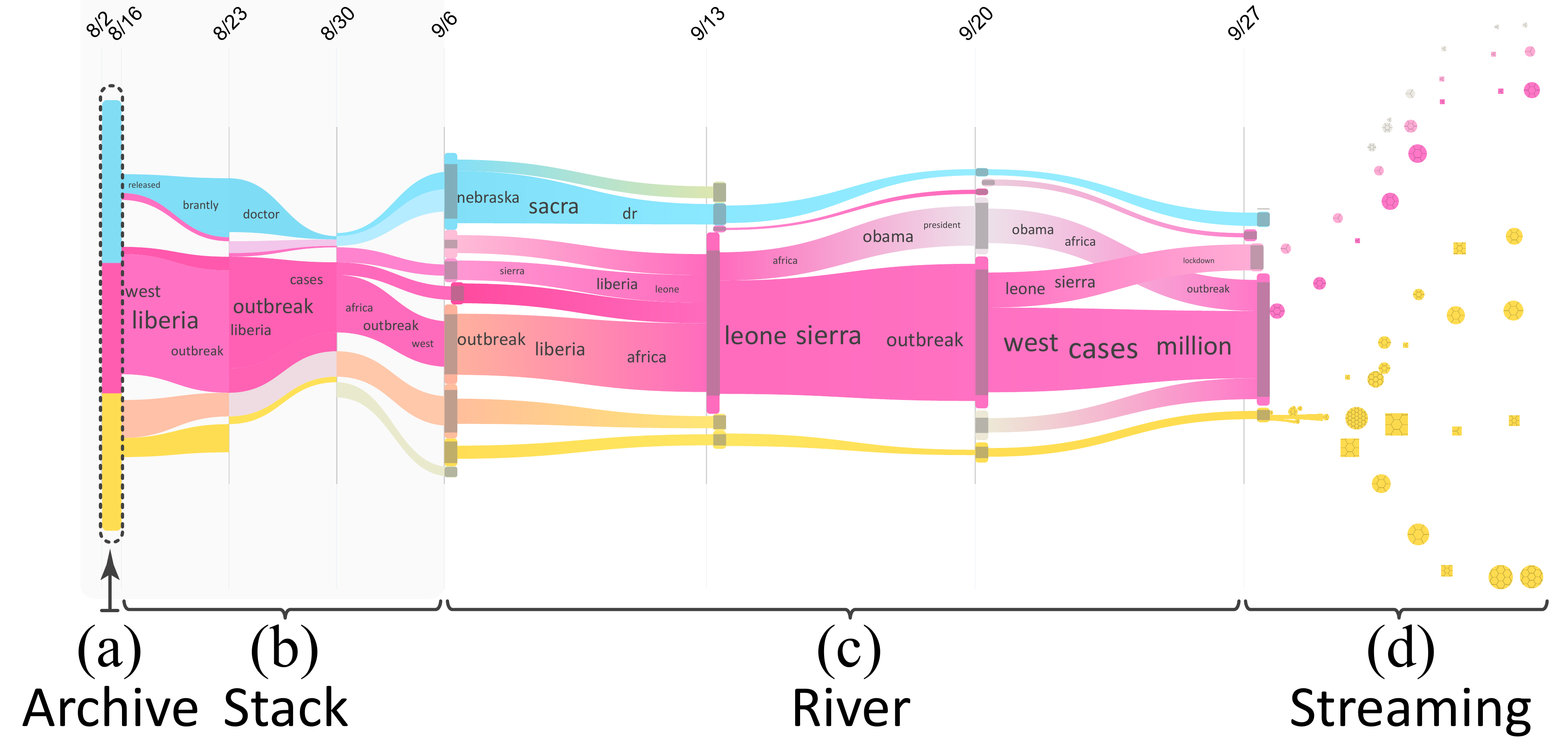}
	\vspace{-4mm}
	\caption{The visualization is divided into four areas: (a) archive; (b) stack; (c) river; (d) streaming.}
	\label{fig:fourAreas}
\end{figure}

\subsection{Visualization Overview}

Based on the guidelines described in Sec.~\ref{sec:rationale}, we \dc{designed} the \emph{\normalsize TopicStream} visualization (Fig.~\ref{fig:fourAreas}).
The \emph{\normalsize x}-axis represents time.
The cut nodes are visualized as vertical bars at the corresponding time step.
The evolutionary relationship between cut nodes is represented by the stripes between the corresponding vertical bars.
The flowing dots on the right side \kg{represent} the newly arrived documents that are currently streaming in.
The different colors encode \kg{various} topics.\looseness=-1

The core \dc{intent} of our visualization is to help users track the dynamic characteristics of text streams.
Every detail in our design \dc{was} carefully crafted to cater \kg{to} this purpose.
For example, sedimentation animation is used to merge newly arrived documents \kg{in} the dominant center of \dc{the} visualization (\textbf{\normalsize R2}).
\kg{As the number of arriving documents increases}, topic bars gradually move to the other side of the display and leave \kg{a} space for new topics (\textbf{\normalsize R1}).
With \kg{such} mechanisms, users can focus on the latest development of topics and \kg{identify} interesting patterns to conduct further analysis.
\kg{In particular,} the visualization consists of four regions (Fig.~\ref{fig:fourAreas}, \textbf{\normalsize R1}):\looseness=-1
\begin{enumerate}

\item \textbf{\normalsize Streaming}, which is on the rightmost side of \dc{the} visualization\kg{,} consists of newly streamed-in documents (e.g., the time period after Sep. 27  in Fig.~\ref{fig:fourAreas}(d)).

\item \textbf{\normalsize River}, which is the dominant region of \dc{the} visualization\kg{,} consists of recent topics \kg{along with} their splitting and merging relationships (e.g., Sep. 6 - 27 in Fig.~\ref{fig:fourAreas}(c)).

\item \textbf{\normalsize Stack}, which is to the left of the river region\kg{,} contains older topics and documents (e.g., Aug. 16 - Sep. 6 in Fig.~\ref{fig:fourAreas}(b)).
To reduce the visual complexity caused by the splitting and merging relationships, this region removes splitting/merging branches and only displays the mainstream of each topic.
Since users want to keep track of how the topics in this region connected with the topics in the river region, the white spaces between the topic stripes are not removed.
The width of each time step in this region is smaller than that in the river region to save space.\looseness=-1

\item \textbf{\normalsize Archive}, which is on the leftmost side, contains the oldest topics and documents (e.g., Aug. 2 - 16 in Fig.~\ref{fig:fourAreas}(a)).
Although the stacked region can \docpr{reduce the} amount of space \docpr{required}, it is still cluttered for a text stream with tens or even hundreds of time steps.
To solve this issue, we introduce the archive region,
which uses a stacked bar (Fig.~\ref{fig:fourAreas}(a)) to represent documents whose times are \emph{\normalsize k} time steps earlier than the newly streamed-in ones.
In \emph{\normalsize TopicStream}, \emph{\normalsize k} is specified by the user.
For example, \emph{\normalsize k} is set to 8 in the example of Fig.~\ref{fig:fourAreas}.
To save space, the width of the bar is fixed no matter how many documents are archived.
Each bar item represents a topic.
Its height represents the average number of documents of each time step that belongs to this region.



\end{enumerate}
As described above, the visualization designs \docpr{for} a bar and a stacked graph are quite straightforward.
We \docpr{will next} introduce the visualization \kg{designs} of the river and streaming regions in detail.


\subsection{Visualization Design}
\subsubsection{Tree Cut as \kg{a} River}

\noindent\textbf{\normalsize Visual Encoding.}
Each cut node is represented by a vertical bar (topic bar) \kg{similar to that presented in~\cite{cui2014}.}
The tree depth of a cut node is represented by the horizontal offset to the time step.
\kg{When} a node in the tree \kg{is deep}, the corresponding topic bar \kg{moves} to the right.

The number of documents contained \kg{in} a topic node is represented by the height of the topic bar.
The width of the colored stripe between two topic bars indicates the number of document pairs between the two bars.
For example, the left width of the stripe represents the portion of documents mapped to the documents in the right topic bar.
The dark region in the middle of a topic bar represents the portion of documents mapped to the documents both in the previous and \kg{the} next topic trees (Fig.~\ref{fig:fourAreas}).

\noindent\textbf{\normalsize Layout.}
The basic representation of the visualization is a directed acyclic graph (DAG).
A node represents a topic and an edge between nodes encodes the evolutionary relationships between topics with mapping.
When a new batch of documents \kg{is} processed, we first run the DAG layout algorithm to \kg{determine} an optimal order for the new topic nodes.
Once the topological structure is computed,  a force model is built to generate the sedimentation animation and merge new documents with existing topic bars.

\begin{figure}[t]
  \centering
 \includegraphics[width=\columnwidth]{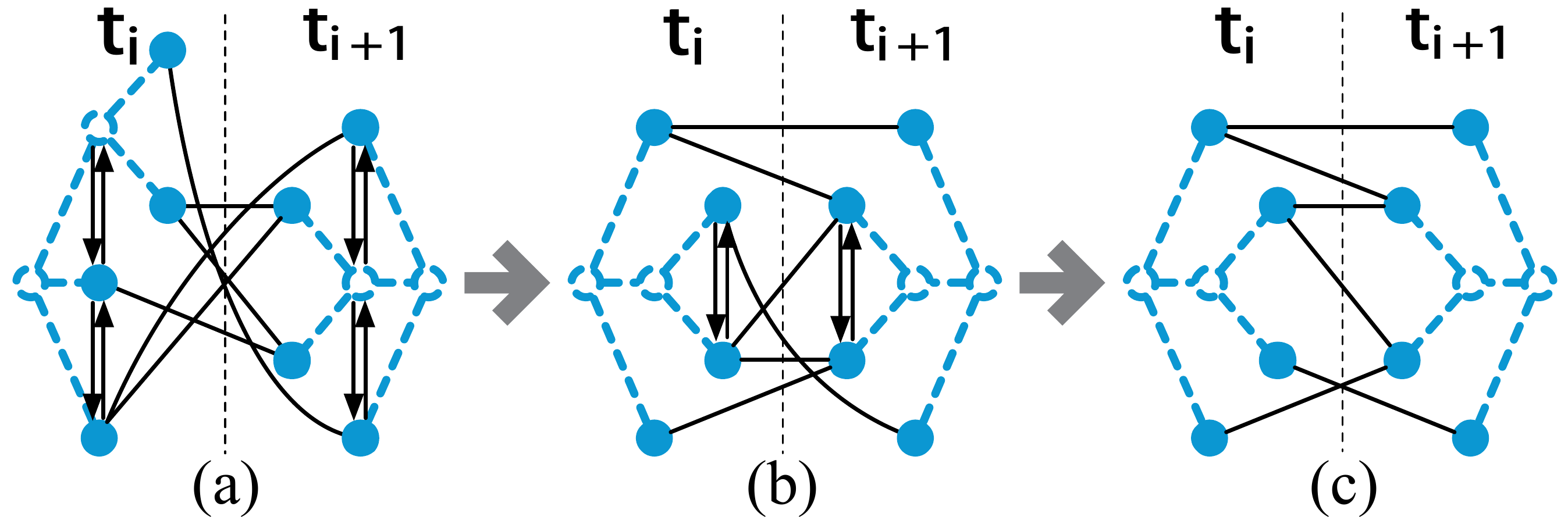}
 \vspace{-3mm}
  \caption{
Reordering example: (a) reorder level one; (b) reorder level two; (c) result.
}
  \label{fig:reorder}
  \vspace{-1mm}
\end{figure}

\begin{figure}[t]
  \centering
 \includegraphics[width=3in]{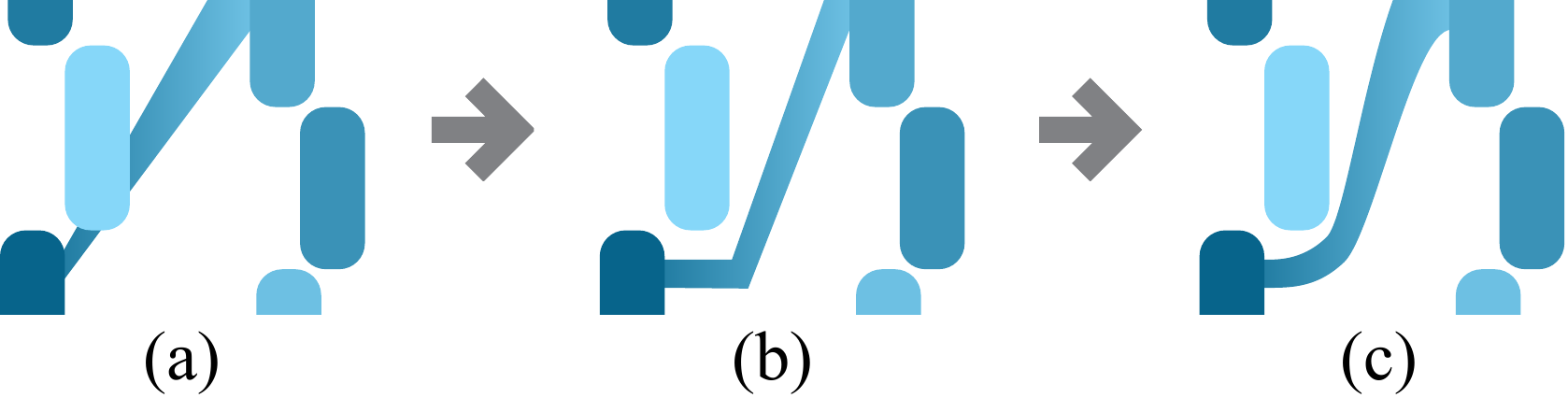}
  \caption{
  Example of edge routing: (a) the stripe is hidden by the topic bar; (b) two intermediate points are added; (c) a B\'{e}zier curve is utilized to improve visual quality.
  }
  \label{fig:route}
  \vspace{-5mm}
\end{figure}


We \kg{initially} reorder the cut nodes at each time step to minimize edge crossings between neighboring time steps \kg{and generate a legible layout that illustrates the evolving patterns.} \kg{Edges are then routed} to avoid overlapping between nodes and edges. \kg{Finally}, representative documents \kg{are packed} on a selected stripe.

\emph{\normalsize Reordering.} Sugiyama's heuristics~\cite{Sugiyama1981}, \kg{which is} a well-known DAG layout algorithm, is employed to reorder the nodes at each time step to minimize edge crossings.
However, if we directly run the algorithm without constraints, sibling nodes \kg{can} be separated by other nodes.
We implement Sugiyama's heuristics from \dc{the} \kg{highest} to the lowest \kg{levels} of the tree at each time \kg{to ensure that the sibling nodes stay together}\dc{.}
Fig.~\ref{fig:reorder} provides an example generated by the reordering algorithm.

\emph{\normalsize Edge Routing.} \kg{Stripes} and topic bars \kg{can} overlap because topic nodes are offset to encode their depth (Fig.~\ref{fig:route}(a)).
We employ the edge routing technique~\cite{Cui2008} to solve this problem.
Two additional intermediate points are introduced for each overlapping part to route the stripe.
The B\'{e}zier curve is utilized to help users follow the striped path (Fig.~\ref{fig:route}).

\begin{figure}[t ]
  \centering
  \vspace{-3mm}
 \includegraphics[width=0.7\linewidth]{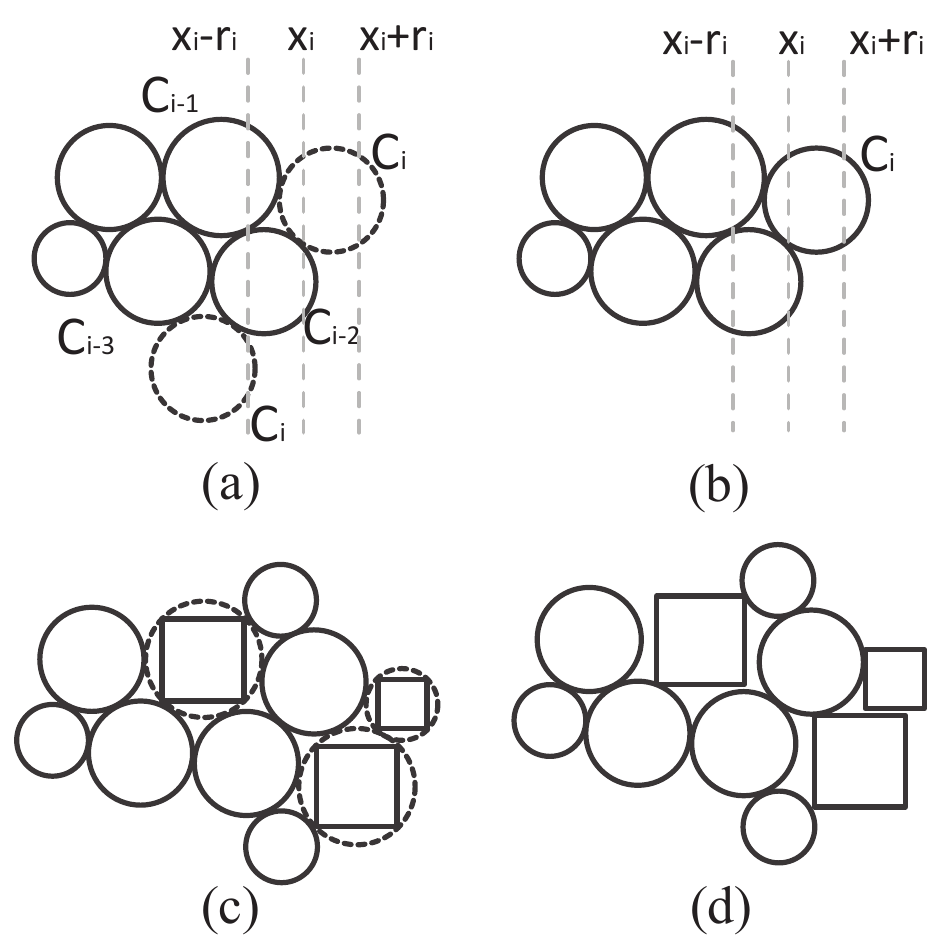}
 \vspace{-1mm}
  \caption{
  Illustration of the packing algorithm:
  (a) \kg{finding} possible placement positions of $C_i$;
  (b) \kg{setting} the position closest to ($x_i$, 0) as the placement position;
  (c) \kg{replacing several} circles \kg{with} the corresponding squares;
  (d) \kg{reducing} the gap with the size \docpr{constraints} and \kg{deriving} the final packing result.\looseness=-1
  }
  \label{fig:packing}
  \vspace{-3mm}
\end{figure}

\begin{figure}[b]
  \centering
  \vspace{-3mm}
 \includegraphics[width=2.6in]{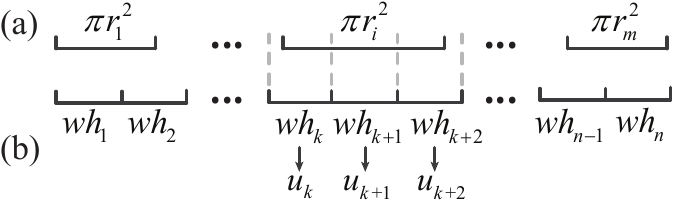}
  \caption{
  \kg{Deriving} the initial \emph{x} position:
  (a) align all the circles on a straight line based on their areas;
  (b) align all the stripe segments on a straight line based on their areas.
  The dotted vertical lines indicate the overlapping relationship \docpr{between} the area of the circle and that of the segmented stripes.
  Based on \docpr{this} relationship, $\normalsize x_i$ is approximated \docpr{as an} $\normalsize average(u_k, u_{k+1}, u_{k+2})$.
  }
  \label{fig:xvalue}
\end{figure}


\emph{\normalsize Packing}. We pack \kg{the documents} on the topic stripe 
(\textbf{\normalsize R3}) \kg{to help users understand and compare their relationships, \docpr{including the incoming} order and similarity relationships.}
Each news article is represented by a circle \kg{in our visualization}, \kg{whereas} each tweet is represented by a square.
For \dc{the sake of} simplicity, each square is approximately represented by a circle whose center is \dc{the same as} the square\dc{'s} and whose radius is $\beta\cdot\sqrt{2}b$.
$\normalsize b$ is the side length of the square and $\beta$ ($1/\sqrt{2}\leq\beta\leq1$) is a parameter \kg{that balances} \dc{the} intersection and gap\kg{s} between elements (e.g., circles and squares) in the final packing result.
The larger $\beta$ is, the \dc{larger the} gap might be.
The packing problem is formulated as a circle packing problem \kg{using this approximation}.
We then employ a front-chain-based circle packing \docpr{algorithm, as in~\cite{Wang2006visualization,ZhaoTVCG2014}, to} pack circles \kg{tightly} on the selected stripe.
Fig.~\ref{fig:packing} illustrate\kg{s} the basic idea of this packing algorithm.

Compared with the packing problem described in~\cite{ZhaoTVCG2014}, our \kg{problem} does not provide the initial \emph{x} coordinate for each circle.
Only the incoming order of each circle is provided in our packing problem.
Thus, we \kg{have} to derive the initial \emph{x} coordinate based on the order of \kg{the} circles.
The basic idea is to \kg{determine} an approximate placement position \kg{for} each circle, which is achieved by approximately mapping its area to the area of the segmented stripes.
The average \dc{of the} \emph{x} coordinates of the corresponding segmented stripes is \kg{then} used to approximate the initial \emph{x} coordinate of the circle.
\kg{In particular}, we align all the circles on a straight line based on their areas (Fig.~\ref{fig:xvalue}(a)).
The area of circle $C_i$ is $\pi{r_i}^2$.
We \kg{then} divide the stripe into \emph{\normalsize n} uniform segments along its \emph{x}-axis.
The height of the \emph{\normalsize k}-th segment is denoted as $h_k$ and its area is $wh_k$,
\kg{where} $w$ is the width of each segment along the \emph{x}-axis.
All these segments are also aligned on a straight line based on their areas (Fig.~\ref{fig:xvalue}(b)).
Fig.~\ref{fig:xvalue} \kg{shows that} the overlapping relationship between the area of the circle and that of the segmented stripes can be \kg{determined using two straight lines}.
For example, the initial $x_i$ of circle \emph{\normalsize i} in this figure is approximated by $\normalsize average(u_k, u_{k+1}, u_{k+2})$.
Here $u_k$ is the \emph{x} coordinate of the center of \emph{\normalsize k}-th segment.

\noindent \textbf{\normalsize Interaction}.
We also provide the following interactions to explore the complex evolutionary clustering results from multiple perspectives \kg{\docpr{in addition to} the interactions described in~\cite{cui2014} (e.g. details on demand, collapsing/expanding time steps, splitting/merging topic bars, and changing focus).}

\emph{\normalsize Document Query}.
Once the documents \kg{are} transformed into a colored stripe, we adopt \kg{circle packing} to encode the documents \kg{contained} within the color stripe for further query and analysis.
The example in  Fig.~\ref{fig:vtreemap} \dc{shows how} users can click the stripe and turn it into a circle/square packing, in which a circle represents a news article \dc{and a} square encodes a tweet.
Once the packing result is displayed, users can manually click one or more \dc{documents} to examine \dc{the} content in detail.

\emph{\normalsize Visual Comparison}.
We allow users to compare the relationships among different time steps by leveraging \dc{a} circle packing algorithm.
For example, users can compare the incoming order and similarity relationships\docpr{, as shown in} Fig.~\ref{fig:ebola}(a).
One of our experts, P2, commented \kg{that}, ``Comparing the incoming order of documents \dc{helps} me easily discover who \kg{talked about} a topic first (\kg{that is, who} set the agenda) and who \docpr{immediately follow\kg{ed}}.
This \kg{feature} \dc{can help me} \kg{study} agenda setting in my field.''

\subsubsection{Streaming Document as Sedimentation}


\begin{figure}[b]
	\vspace{-3mm}
	\centering
	\includegraphics[width=\columnwidth]{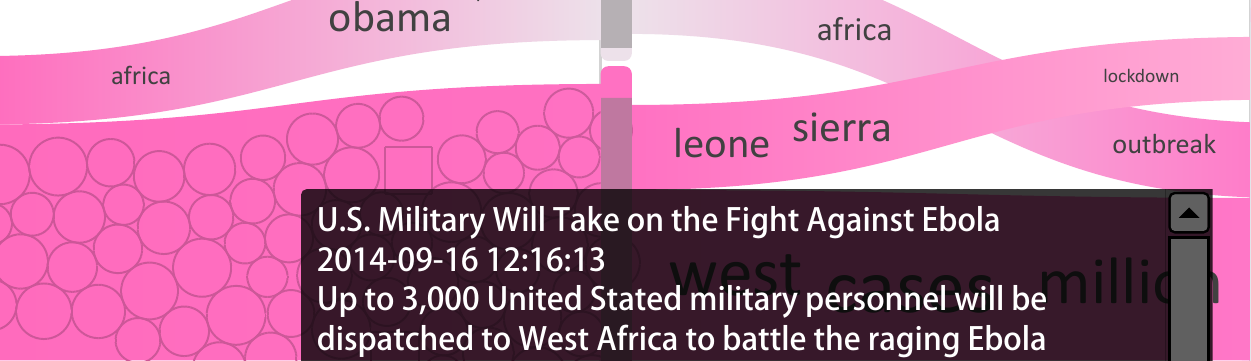}
	\vspace{-5mm}
	\caption{
	Encode documents after sedimentation \docpr{as \kg{circle/square packing}}.
	}
	\label{fig:vtreemap}
\end{figure}

\noindent \textbf{\normalsize Visual Encoding}.
Inspired by visual sedimentation~\cite{Huron2013visual}, we use the river sedimentation metaphor to encode the process of newly arrived text documents \kg{that merge} with existing topics (\textbf{\normalsize R2}).
To quicken the sedimentation process of a high-volume text stream, a set of document clusters are derived from the incoming documents by using k-means clustering.
A token is a visual mark representing a document cluster.
The generation process of the sedimentation metaphor consists \dc{of} four steps:

\emph{\normalsize Entrance}.
Newly arrived documents are represented as circular or rectangular tokens (Fig.~\ref{fig:fourAreas}) \kg{that come} into view from the right side.
Documents \kg{with} similar content are clustered into one token, the size of which indicates the \kg{number of} documents, \kg{to handle the scalability issue}.
The color of each token encodes the topic that it contains.

\emph{\normalsize Suspension}.
Each token \kg{moves toward} (from right to left) the corresponding topic bars of the latest time step.
Token \dc{size} \dc{decreases} gradually \kg{during the movement}.

\emph{\normalsize Accumulation} and \emph{\normalsize decay}.
\kg{The tokens} will stop moving and start \kg{to} decay \kg{once they touch the corresponding topic bars or other tokens that have already settled.}
The settled tokens continue to shrink and merge \kg{with} existing topics.

\emph{\normalsize Aggradation}.
The colored stripes continue to grow and \kg{indicate} the latest development of topics \kg{when the settled tokens are resolved}.

Once a batch of documents (e.g., for a day) \docpr{are sedimented}, \kg{the} corresponding topic bars appear and push older topic bars to the \dc{left-hand} side.
The archive and stack regions \kg{then} change accordingly.

\begin{figure}[t]
	\centering
	\includegraphics[width=\columnwidth]{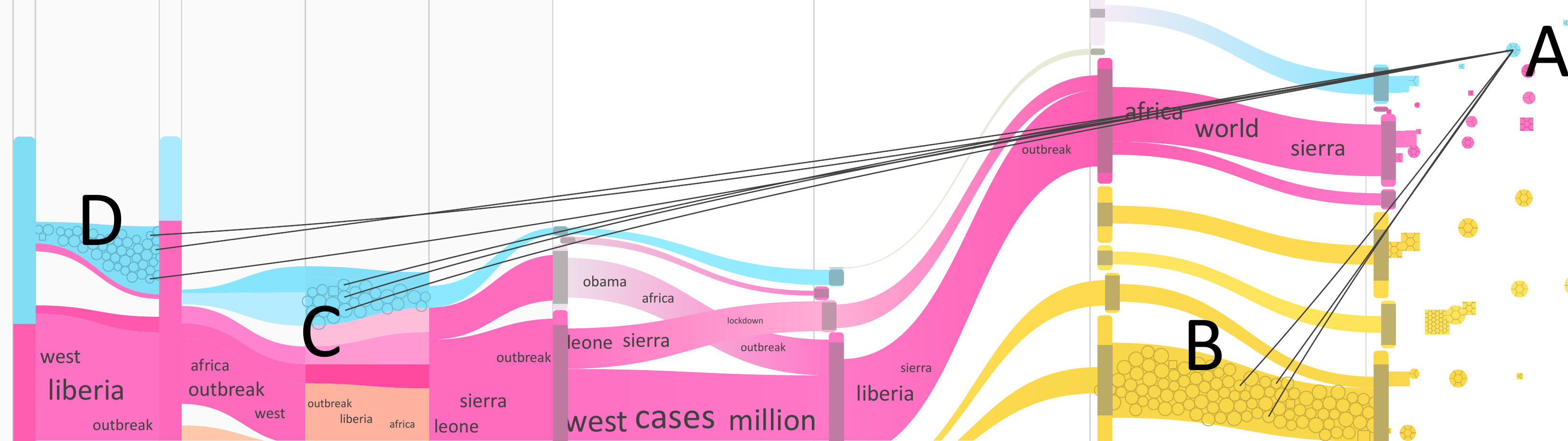}
	\vspace{-4mm}
	\caption{
	Relevant documents of cluster A are highlighted \docpr{in the} river (B)\xiting{, the stack (C),} and the archive (\xiting{D}) regions.\looseness=-1
	}
\vspace{-5mm}
	\label{fig:dochighlight}
\end{figure}

\noindent \textbf{\normalsize Layout}.
Each token is \kg{assigned to} a region based on the topological structure in the ``reordering and edge routing'' step \kg{during the sedimentation process.}
The token can only move within the \dc{assigned region} and cannot cross the border.
The \dc{speed} of the token is controlled by two forces: 1) a universal gravity force and 2) an attractive force between the token and the \kg{sedimented} tokens.
The gravity force provides each token \kg{with} constant acceleration from right to left.
The attractive force ensures that similar documents will sediment close to \kg{one another.}
Therefore, the total acceleration $a_k$ for a moving token $k$ is defined as
$$a_k=g+\sum_is_{ik}*{n_i}/{||p_i-p_k||^2},$$
where $g$ is the constant gravity acceleration, $p_i$ is the location of \kg{sedimented} token $i$, $p_k$ is the location of token $k$, $n_i$ is the number of documents in token $i$, and $s_{ik}$ is the content similarity between token\kg{s} $i$ and $k$.

\noindent \textbf{\normalsize \xiting{Interaction.}}
The sedimentation visualization also allow\kg{s} users to examine the content of \docpr{the} \kg{incoming} documents \kg{interactively} and compare them with older documents.

\emph{\normalsize Document Link}.
In many text stream analysis tasks, it is \dc{desirable} to quickly find related documents covering a long \dc{time} period.
Document link is supported for this \kg{requirement of our system}.
For example, users \kg{can initially} explore the content in the streaming region and find a document/cluster of interest.
Our system \kg{then} automatically use\kg{s} the word vector in the given document and locate\kg{s} the most similar documents in all three regions (i.e., streaming, stack, and archive).
Once the related documents/clusters are located, the connections \kg{are} displayed for users to \kg{explore further}.

An example of document \dc{link} is shown in Fig.~\ref{fig:dochighlight},
\kg{in which} a user explores relevant documents \kg{from an incoming Twitter} cluster (Fig.~\ref{fig:dochighlight}A).
Relevant documents are found in the river (Fig.~\ref{fig:dochighlight}B), stack (Fig.~\ref{fig:dochighlight}C), \dc{and} archive (Fig.~\ref{fig:dochighlight}D) \kg{regions}.
The archive region is expanded accordingly \kg{to facilitate the examination of the relevant documents.}

\kg{Users} can also click on a token while it is still in the suspension step.
\dc{Related} documents \kg{are} \dc{then} displayed for further examination.

%

\section{Quantitative Evaluation}
\label{sec:quantitativeevaluation}
In this section, a quantitative evaluation of the proposed streaming tree cut algorithm is conducted.


\subsection{Fitness and Smoothness}
To assess the effectiveness of \kg{the} streaming tree cut algorithm, we \tvcgminor{compared our algorithm with a baseline algorithm in terms of fitness and smoothness.}

\subsubsection{Criteria}
Fitness and smoothness \kg{are} two important criteria to evaluate the derived streaming tree cuts.
\dc{Fitness} measures how satisfactorily the topics on the tree cut represent the topic distribution within a topic tree.

\noindent\textbf{\normalsize Fitness ($\bm{F}$)}:
We derived the measure from the proposed tree cut likelihood equation, $\bm{F}=p({\phi}^t|T^t)p(\mathcal{D}_{f}|{\phi}^t)$,
where the right side is defined in Eq.~(\ref{eq:hmm2}).
$p({\phi}^t|T^t)$ describes how the tree cut \xiting{fits} the tree and $p(\mathcal{D}_{f}|{\phi}^t)$ describes how it \xiting{fits} the focus.
A larger  $F$ value \kg{indicates a} better tree cut.

The following three measures \kg{assess} the smoothness between \kg{the} adjacent tree cuts.
In \kg{the} implementation, \kg{a} larger smoothness value \kg{means that} the two adjacent tree cuts \kg{are smoother}.

%
%
%

\noindent\textbf{\normalsize Tree mapping ($\bm{S_{map}}$)}:
\kg{The} measure \kg{is derived} from the smoothness cost function of the streaming tree cut algorithm, $\bm{S_{map}}(\phi^t,\phi^{t-1})=-E_2{(\phi^t,\phi^{t-1})}$,
where $E_2{(\phi^t,\phi^{t-1})}$ is defined in Eq.~(\ref{eq:hmm5}).


\noindent\textbf{\normalsize Normalized Mutual Information (NMI) ($\bm{S_{NMI}}$)}:
The NMI measure \kg{represents} the mutual information \docpr{shared by both} the cluster assignments and a pre-existing \kg{label}.
\dc{The} Hungarian algorithm~\cite{Steiglitz1982} \kg{is} employed to find the optimal match between the document sets of the two tree cuts.
This measure \kg{assesses} the similarity between adjacent tree cuts.

\noindent\textbf{\normalsize Tree distance ($\bm{S_{dist}}$)}:
\kg{This} measure \kg{is used} to evaluate the difference between \kg{the} tree cuts by aggregating the tree distance between two related cut nodes \xiting{$T_s$ and $T_r$}, 
\xiting{
\begin{eqnarray}
\small
S_{dist}(\phi^t,\phi^{k}) =  -\left( Avg_{T_r,T_s\in \mathcal{C}_{\phi^t}}(D_{T^t}(T_r,T_s)-D_{T^k}(T_r,T_s))^2\right. \nonumber\\
+\left. Avg_{T_r,T_s\in \mathcal{C}_{\phi^k}}(D_{T^k}(T_r,T_s)-D_{T^t}(T_r,T_s))^2\right) /2,
\label{eq:distancemesaure}
\vspace{-1mm}
\end{eqnarray}
}
where $D_T(T_r,T_s)$ is the tree distance \xiting{between $T_r$ and $T_s$ under $T$.
If $T_r$ and $T_s$ are not in $T$, they are mapped to $T$}.


\subsubsection{Experimental Settings}
\kg{A} baseline system \kg{was implemented} according to the DOI-based tree cut generation method~\cite{cui2014}.
To compare the fitness and smoothness of \kg{the proposed} methods to the baseline, we conducted experiments on the following two datasets.

\begin{compactitem}
\item \textbf{\normalsize Dataset \emph{A}} \kg{contains} 207,406 news articles and 15,565,532 tweets related to ``Ebola'' (\kg{from} Jul. 27, 2014 to Feb. 21, 2015).
The articles were organized into 30 topic trees by week.
The tree depth, total node number, and first-level node number of the trees varied from 3 to 5, 34 to 223, and 10 to 33, respectively.
\item \textbf{\normalsize Dataset \emph{B}} \kg{contains} 543,114 news articles related to ``Obama''  (\kg{from} Oct. 14, 2012 to Feb. 21, 2015).
The articles were organized into 62 topic trees by every \kg{two} weeks.
The tree depths varied from 4 to 11, the total node numbers changed from 246 to 471, and the node number of the first level ranged from 18 to 79.\looseness=-1
\end{compactitem}


\begin{table*}[t]

\vspace{-3mm}
    \caption{
    Evaluation of the overall likelihood and smoothness.}
    \vspace{-3mm}
    $f_r(\cdot)=\frac{m_o-m_b}{m_b}*100\%$, where $m_b$ and $m_o$ are the measure values of the baseline method and our method.
    \label{table:average}
    \vspace{1mm}
    \centering \scalebox{0.8}{
    \begin{tabular}{|c|c|c|c|c|c|c|c|c|c|}

    \hline
    \multirow{2}{*}{Dataset}&{$f_r(\bm{F})$(\%)} & {$f_r(\bm{S_{map}})$(\%)}& \multicolumn{3}{c|}{$f_r(\bm{S_{NMI}})$(\%)} & \multicolumn{3}{c|}{$f_r(\bm{S_{dist}})$(\%)}\\
    \cline{2-9}
    &$F({\phi}^{t})$ & $S_{map}({\phi}^{t},{\phi}^{t-1})$ & $S_{NMI}({\phi}^{t},{\phi}^{t-1})$ & $S_{NMI}({\phi}^{t},{\phi}^{t-2})$ & $S_{NMI}({\phi}^{t},{\phi}^{t-3})$& $S_{dist}({\phi}^{t},{\phi}^{t-1})$ & $S_{dist}({\phi}^{t},{\phi}^{t-2})$&$S_{dist}({\phi}^{t},{\phi}^{t-3})$\\

    \hline
   \emph{A} & 4.5486 & 18.7873 & 7.6839 & -1.8875 & -2.6680 & 13.5781 & -1.5226 & -2.4843 \\
    \emph{B} & 7.5084 & 26.4451 & 8.7131 & -3.1711 & -3.2452 & 13.1334 & -4.2689 & -5.4008 \\
        \hline
    \end{tabular}
    }
    \vspace{-5mm}
\end{table*}

\begin{figure}[t]
  \vspace{-1mm}
  \centering
  \includegraphics[width=0.7\columnwidth]{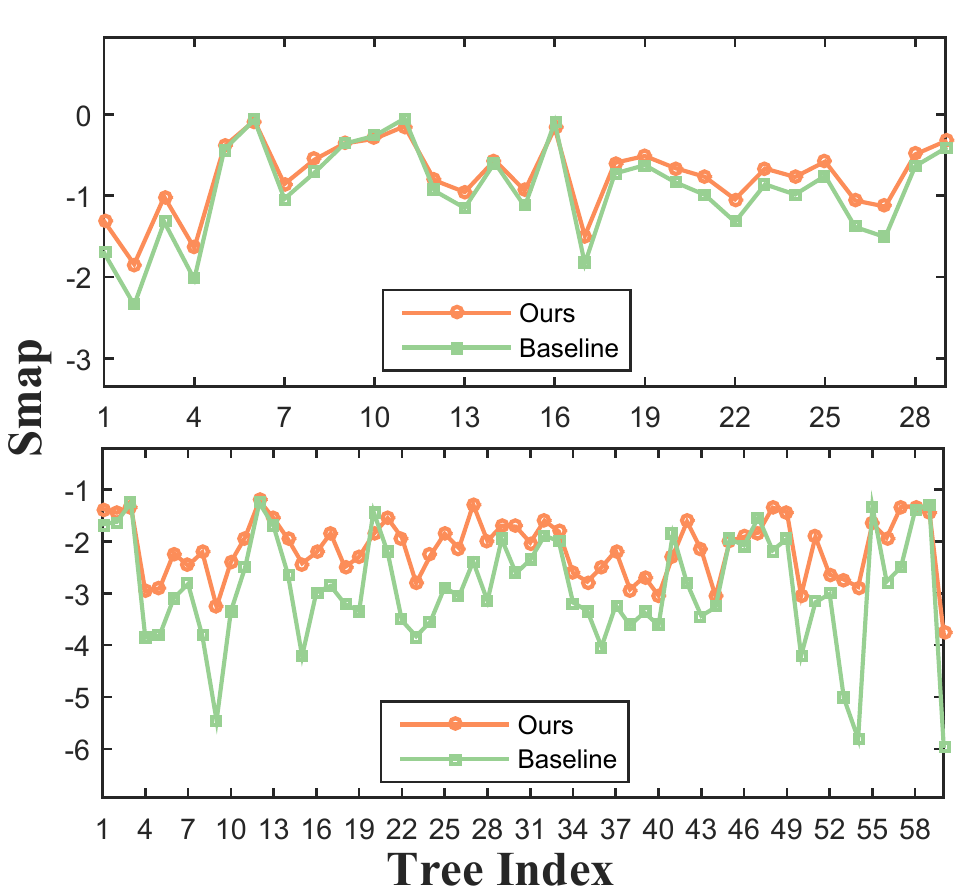}
  \vspace{-1mm}
  \caption{
  Comparison of tree mapping smoothness.}
  \vspace{-5mm}
  \label{fig:smap}
\end{figure}

\begin{figure*}[b]
	\centering
\vspace{-5mm}
    \includegraphics[width=\linewidth]{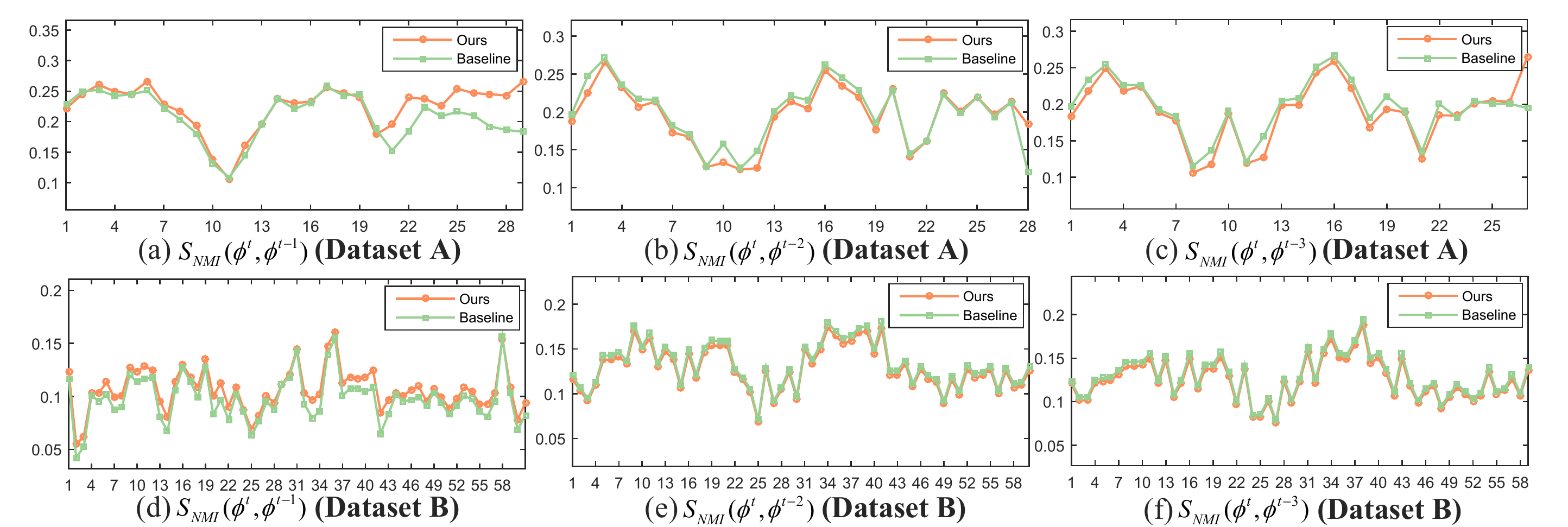}
\vspace{-5mm}
	\caption{
Comparison of NMI smoothness. \kg{X}-axis represents tree index and \kg{Y}-axis encodes NMI smoothness.
}\label{fig:NMI}
\vspace{-3mm}
\end{figure*}

\begin{figure*}[b]
	\centering
    \includegraphics[width=\linewidth]{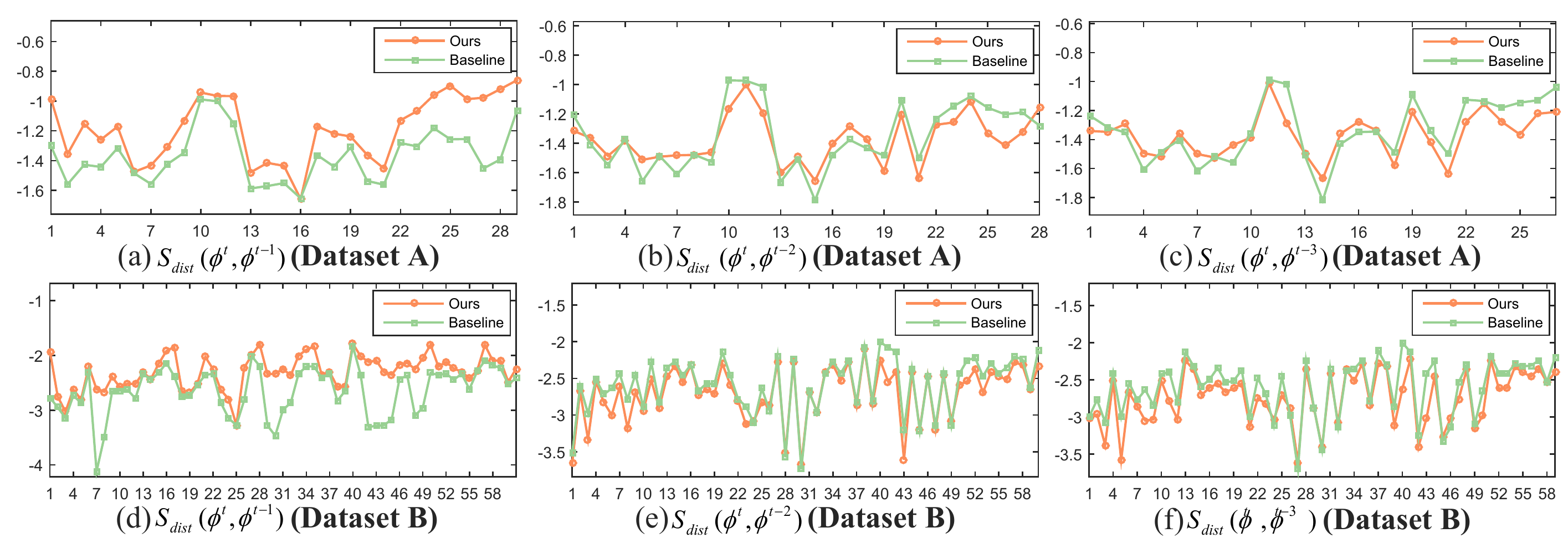}
\vspace{-5mm}
	\caption{
Comparison of tree distance smoothness. \kg{Y}-axis encodes tree distance smoothness.\looseness=-1
}\label{fig:distance}
\end{figure*}

To eliminate bias caused by the focus node selection, the same number of focus nodes \kg{was randomly selected 50} times and the experiments \kg{were repeated 50} times.
At each time, ${F}$ for each tree cut \kg{was computed}.
Since the measure ${S_{map}}$ was defined on adjacent tree cuts, we only \dc{computed} ${S_{map}}$ between adjacent tree cuts.
To demonstrate the global smoothness of \kg{the proposed} algorithm, ${S_{NMI}}$ and  ${S_{dist}}$， were computed between ${\phi}^t$ and each of ${\phi}^{t-1}$, ${\phi}^{t-2}$, \kg{and} ${\phi}^{t-3}$.
The results were computed by averaging the 50 trials.

\subsubsection{Results}
\label{sec:results}

\kg{The} overall fitness and smoothness \kg{were compared} with the baseline.
As shown in Table~\ref{table:average}, \kg{the proposed} method \docpr{generates} a much smoother structure than the baseline while maintaining \docpr{greater} fitness.
When the smoothness between non\kg{-}adjacent tree cuts \kg{\docpr{was} compared}, \kg{the proposed} method \kg{performed slightly} worse\kg{,}
because \kg{the} method only \kg{considered} the adjacent tree cuts to improve the performance \dc{of the data stream}.
Thus\kg{,} the global smoothness \kg{was} not maintained to \kg{a certain} extent.

We further compared the smoothness of our method with the baseline between trees under these measures.
As shown in  \dc{Figs}.~\ref{fig:smap}, \ref{fig:NMI}, and \ref{fig:distance}, the proposed streaming algorithm worked as well as the baseline under the three measures for adjacent tree cuts.
For non-adjacent tree cuts, the smoothness of \kg{the proposed} algorithm \dc{was} \kg{slightly} worse under the commonly used measures NMI and tree distance.
The fitness of \kg{the proposed} algorithm at each tree was also evaluated.
As shown in Fig.~\ref{fig:sf}, \kg{the proposed} algorithm \dc{was} \kg{more effective} than the baseline at each time in all the datasets.
\kg{These} findings demonstrate that \kg{the proposed} algorithm \docpr{can} preserve the smoothness between \kg{the} adjacent trees as well as the fitness without sacrificing global smoothness. 

\subsection{Scalability}

\begin{figure}[t]
	\centering
	\includegraphics[width=0.7\columnwidth]{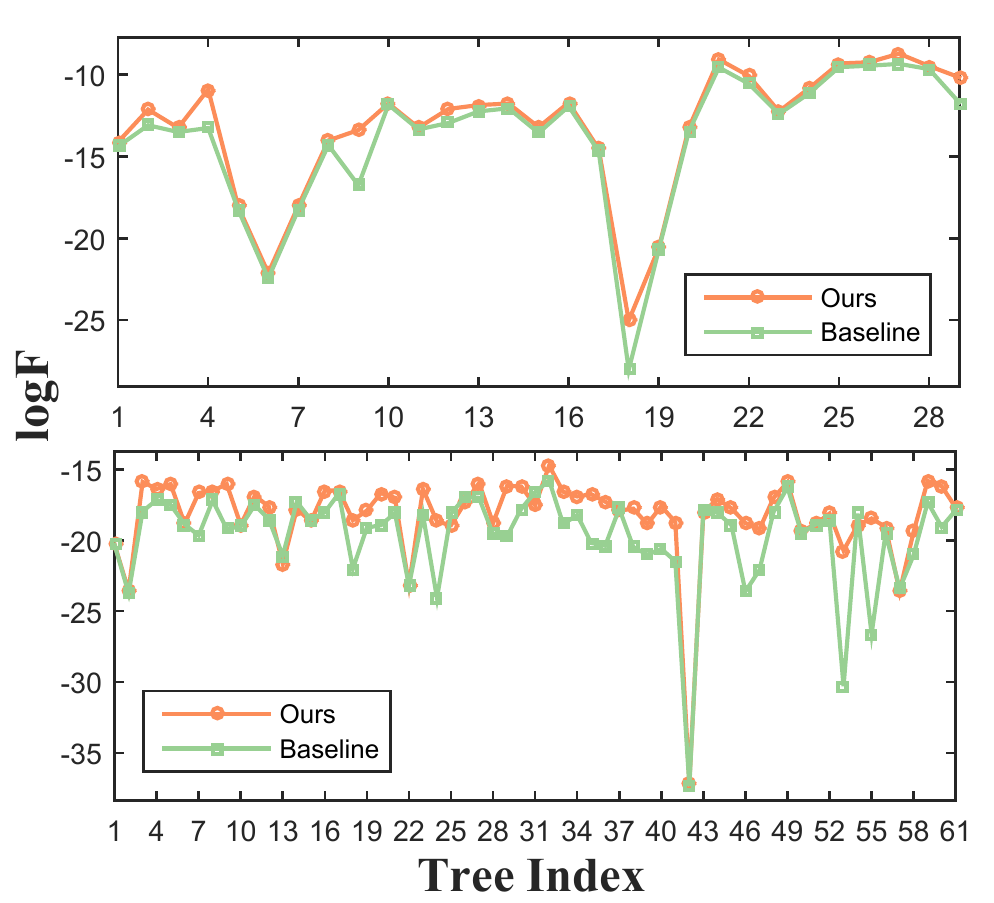}
	\vspace{-1mm}
	\caption{
		Comparison of fitness at each tree.}
	\vspace{-6mm}
	\label{fig:sf}
\end{figure}

We conducted two experiments to evaluate the scalability of our algorithm.
In the first experiment, we investigated the ability of our algorithm to handle topic trees with a large number of internal nodes ($I_{num}$).
In the \docpr{second, we} tested the ability of our algorithm to process long sequences of topic trees.



\subsubsection{Experimental Settings}

The dataset used in the first experiment was generated by copying the first ten trees in Dataset \docpr{A \emph{\normalsize s}} times ($s\in \{1,3,...,15\}$).
As a result, we obtained eight groups of topic trees with varied $I_{num}$ ($I_{num}\in\{118, 354, ..., 1770\}$). 
For each group of topic trees, we treated the first five trees as old trees and evaluated the average time to process the 6th to 10th trees. 
In our experiments, focus nodes were randomly selected to avoid any biased conditions.
To eliminate randomness caused by the focus node selection, we randomly selected the given number $m$ ($m\in\{1,3,5\}$) of focus nodes 50 times and ran the experiment 50 times.
Results were computed by averaging the 50 trials.


In the second experiment, we used the 30 topic trees in Dataset A.
Specifically, we regarded the first $P_{num}$ ($P_{num}\in \{7,9,...29\}$) trees as old trees, and evaluated the time to process the $(P_{num}+1)$-th tree.
\docpr{All other settings were the} same as the first experiment.

The experiments were run on a workstation with an Intel Xeon E5-2630 CPU (2.4 GHz) and 64GB Memory.

\subsubsection{Results}

As shown in Fig.~\ref{fig:exp-time-Inum}, the running time of our algorithm increases at an approximate quadratic rate with the increase of $I_{num}$.
For $m=5$, our algorithm can process topic trees with \docpr{1,770} internal nodes in 66 seconds.
This demonstrates that our algorithm can handle large topic trees.

Next, we \docpr{demonstrated} the scalability of our algorithm in regards to $P_{num}$ under different \emph{\normalsize m}.
\docpr{We} used \docpr{a} normalized running time to eliminate \docpr{any} bias caused by different sizes of the topic trees.
Normalized running time is calculated by multiplying real running time with $(I_{avg}/I_{cur})^2$.
Here $I_{avg}$ is \docpr{computed} by averaging $I_{num}$ of all trees and $I_{cur}$ is $I_{num}$ of the $(P_{num}+1)$-th topic tree.
As shown in Fig.~\ref{fig:exp-time-Pnum}, the normalized running time increases slowly with the increase of $P_{num}$ and the results are consistent across different $m$.
This indicates that our algorithm can process long sequences of topic trees efficiently.\looseness=-1

\begin{figure}[t]
\centering
{
\includegraphics[width=0.36\textwidth]{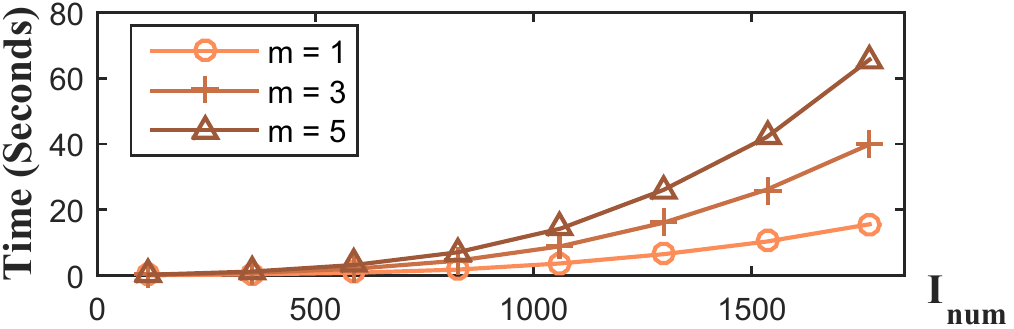}
}
\caption{Running time vs. number of internal nodes in the topic tree ($I_{num}$) vs. number of focus nodes ($m$).}
\label{fig:exp-time-Inum}
\vspace{-1mm}
\end{figure}

\begin{figure}[t]
\centering
{
\includegraphics[width=0.36\textwidth]{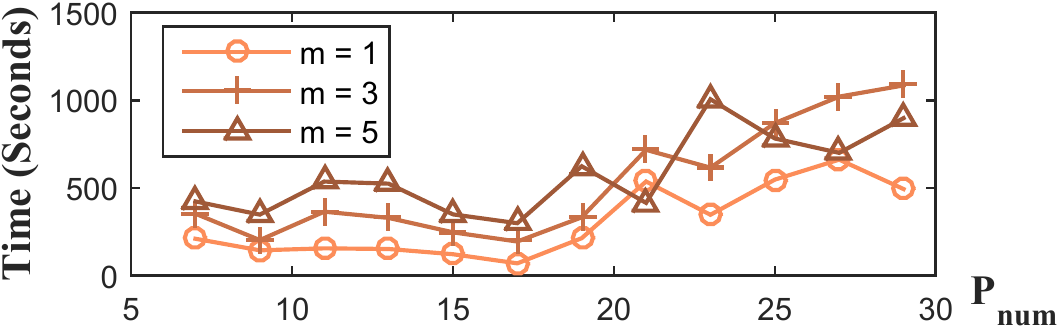}
}
\caption{Normalized running time vs. number of topic trees processed ($P_{num}$) vs. number of focus nodes ($m$).}
\label{fig:exp-time-Pnum}
\vspace{-5mm}
\end{figure}

\section{Case Study}
In this section, we demonstrate the usage scenarios of our approach \dc{using} real-world datasets.

\begin{figure*}[t]
\centering
\centering
  \includegraphics[width=\linewidth]{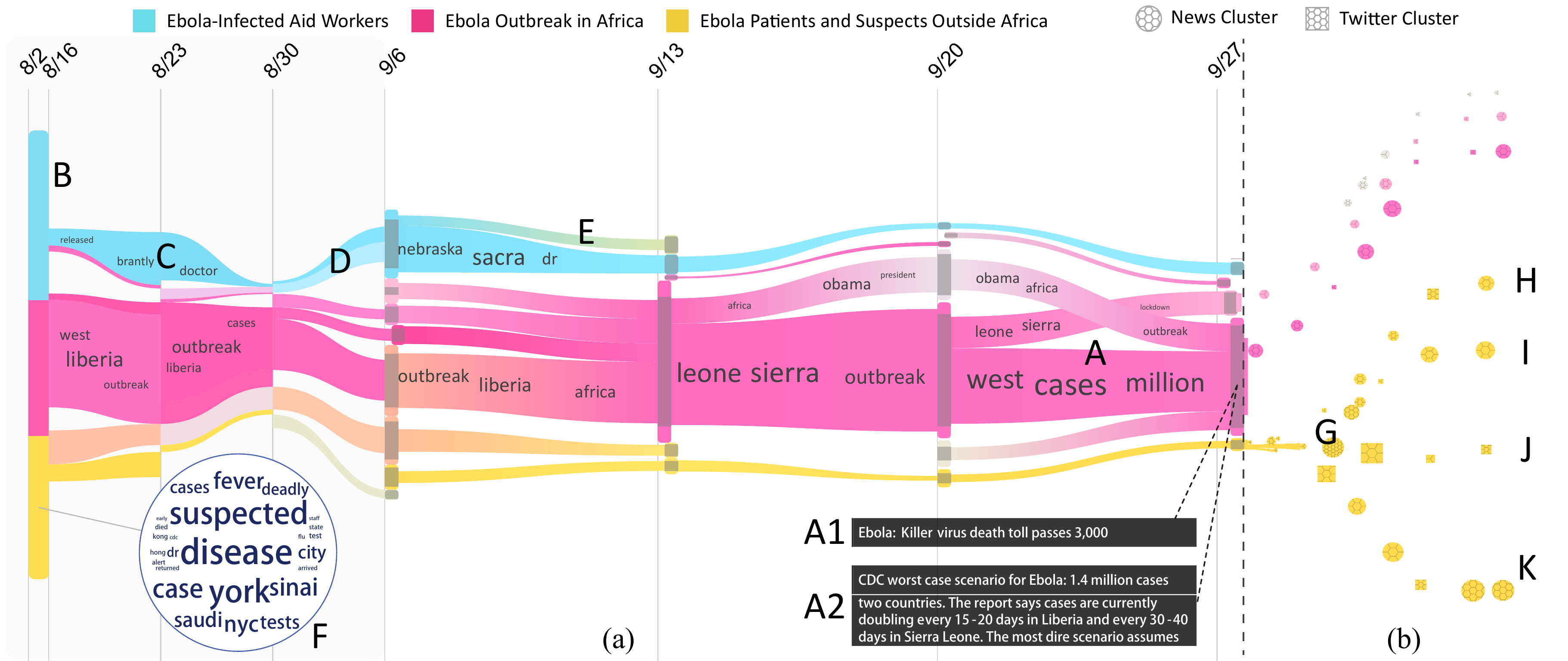}
    \vspace{-3mm}
  \caption{
  Comparative analysis between the severity of \dc{the} epidemic and the intensity of public opinion in the Ebola dataset.
  (a)  \kg{The region with the most severe cases (i.e., Africa)} was the key focus of public opinion \kg{before Sep. 27, 2014}.
  \kg{Epidemic severity during this period} was consistent with \kg{public opinion intensity.}
  (b) An explosive growth of public discussion \docpr{in} non-African regions \kg{occurred after Sep. 27.}
  \kg{Public opinion intensity during this period was inconsistent with epidemic severity.}
  }
\label{fig:msoverview}
\vspace{-5mm}
\end{figure*}
\subsection{Ebola Data}

\kg{The} case study \kg{was conducted} with a professor (P2) \kg{who} majored in public opinion analysis \dc{in} healthcare.
In this case study, we illustrate how TopicStream helps an expert examine the relationship between the severity of an epidemic (e.g., Ebola) and the intensity of public opinion.
The severity of \kg{the} epidemic \docpr{was} measured by the reported \kg{number of cases} and \kg{deaths}.
The intensity of public opinion \docpr{was} represented by the number of news articles and tweets at that time step (the width of the topic stripe).
A wider stripe indicated more intense public opinion (Fig.~\ref{fig:msoverview}).\looseness=-1

A dataset that contains both news articles and tweets collected by using \docpr{the} keyword ``Ebola'' \kg{was used} (Dataset A).
Table~\ref{table:ebola} shows the statistics of the dataset.
\begin{table}[h]

\vspace{-3mm}
    \vspace{1mm}

    \centering
    \scalebox{0.8}{

    \begin{tabular}{|c|c|c|c|c|c|}

    \hline
    Data & Time span & $N_{num}$ & $T_{num}$ & \xiting{$h$} & $I_{num}$ \\
    \hline
   \emph{Old} & 7/27/2014-9/27/2014 & 51,318 & 7,161 & 3-4 & 77-150\\
   \hline
    \emph{New} & 9/28/2014-2/21/2015 & 156,088 & 15,558,371 & 3-5 & 34-223\\
   \hline
    \end{tabular}
    }
    \caption{
\kg{Statistics} of the Ebola dataset. \xiting{Here $N_{num}$ denotes the number of news articles, $T_{num}$ represents the number of tweets, $h$ is the tree depth, and $I_{num}$ denotes the number of internal nodes in the tree.}
}
\vspace{-5mm}
    \label{table:ebola}
\end{table}

\noindent \textbf{\normalsize Spread of Ebola outbreak.}
We first provided the professor (P2) with an overview of the old Ebola data.
The old data (before Sep. 27, 2014) is shown in Fig.~\ref{fig:msoverview}(a).
The news articles from Sep. 28 to Oct. 4 \kg{appeared in a streaming manner, as shown in} Fig.~\ref{fig:msoverview}(b).
\kg{Using} topic keywords and corresponding news articles in Fig.~\ref{fig:msoverview}(a), 
\kg{P2} immediately identified the major topics in the news stream, which \docpr{were} encoded \docpr{as} blue, pink, and \docpr{yellow.}
As in~\cite{cui2014}, we used the mean-shift clustering algorithm to cluster the topic at the first level since it is the most abstract level and can represent the topic tree very well.
For each cluster, we chose the topic closest to the cluster center as one focus topic.

By examining the incoming news articles on the pink topic stripe, ``Ebola outbreak in Africa'' (Fig.~\ref{fig:msoverview}A), \kg{P2} found that the epidemic \kg{was extremely} serious in Africa.
\kg{The epidemic caused a large number of deaths} (Fig.~\ref{fig:msoverview}A1) \kg{and the} \dc{spread} of infections \kg{was rapid}.
For example, \docpr{the} news article \docpr{entitled} ``CDC worst case scenario for Ebola: 1.4 million cases'' mentioned that reported cases in Liberia \docpr{were} doubling every 15 to 20 days and those in Sierra Leone \docpr{were} doubling every 30 to 40 days (Fig.~\ref{fig:msoverview}A2).
The blue topic stripe \kg{contains} keywords ``dr,'' ``sacra,'' \kg{and} talks about ``Ebola-infected aid workers.''
\kg{``Sacra''} is the last name of \kg{Dr. Rick Sacra}, one of the aid workers.
By examining the news articles in the archive area (Fig.~\ref{fig:msoverview}B), \kg{P2} learned that two aid workers returned to the U.S. for \kg{treatment}.
The increased width in the stack area (Fig.~\ref{fig:msoverview}C) \kg{discussed} their recovery.
Figs.~\ref{fig:msoverview}D and \kg{\ref{fig:msoverview}E} in the river area are \dc{related to} the third and fourth infected aid workers.
From the \kg{preceding} exploration, \kg{P2} concluded that several aid workers \kg{had been infected;} however, the situation \kg{was} not serious.
From keywords ``suspected,'' ``york,'' \dc{and} ``sinai'' in the word cloud of the yellow topic stripe (Fig.~\ref{fig:msoverview}F)),  
\kg{P2} concluded \kg{that} this topic \kg{was} about ``Ebola patients and suspects outside Africa.''
After reading the corresponding news articles before Sep. 27, \kg{P2} concluded that only \kg{a few} suspects \kg{were} outside Africa and the situation \kg{was} not serious.\looseness=-1

\begin{figure*}[t]
	\centering
	\vspace{-2mm}
	\centering
	\includegraphics[width=\linewidth]{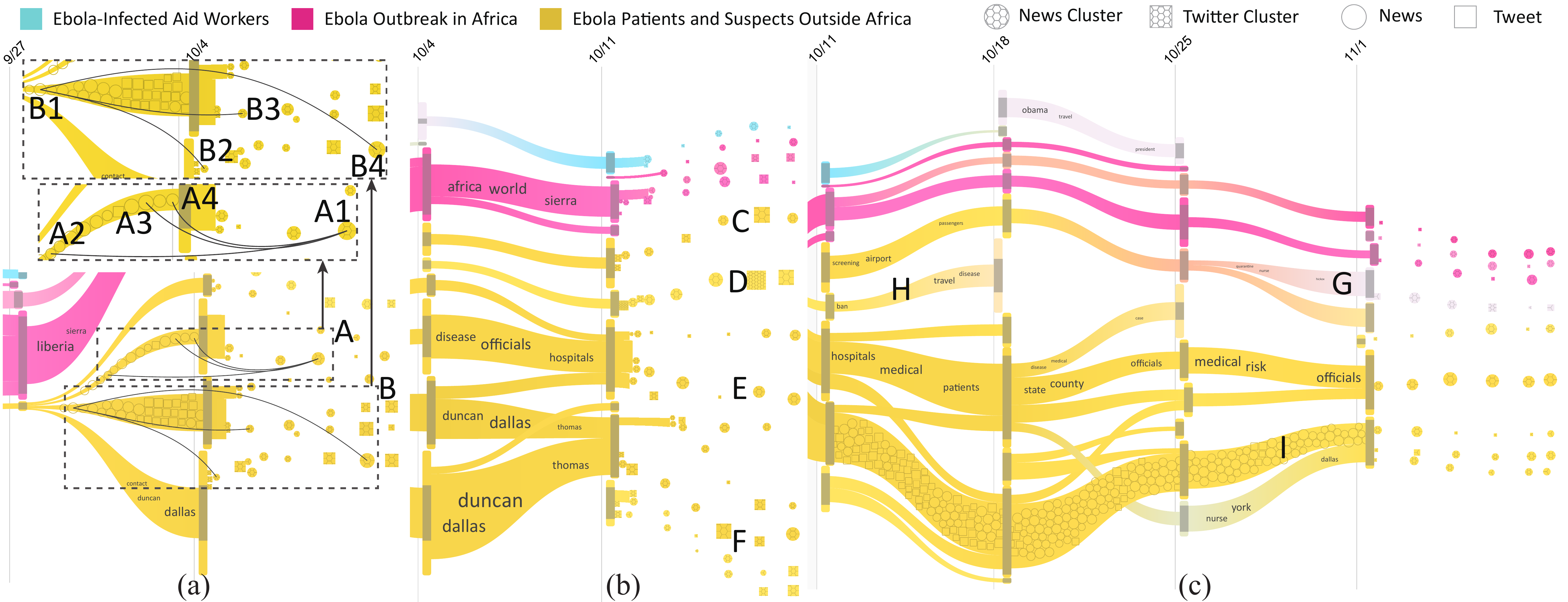}
	\vspace{-3mm}
	\caption{
		Explosive discussion \docpr{of} reported \kg{cases} outside Africa: (a) Oct. 5 to 11; (b) Oct. 12 to 18; (c) Nov. 2 to 8.
	}
	\vspace{-5mm}
	\label{fig:ebola}
\end{figure*}

\noindent \textbf{\normalsize Explosive discussion on Ebola outside Africa.}
\kg{P2} found that the severity of \kg{the} epidemic \kg{was} consistent with the intensity of public opinion before Sep. 27 (Fig.~\ref{fig:msoverview}(a))\kg{,}
that is, the stripe \kg{was wider}, \kg{which indicated} more intense public opinion.
However, as indicated by the increasing number of yellow circles and squares in the visualization (Fig.~\ref{fig:msoverview}(b)), there is an explosive discussion on Ebola outside Africa \kg{occurred} after Sep. 27.
P2 was curious about such a change\kg{;} \kg{thus, the exploration of} the incoming data \kg{continued}.

She noticed that the explosion began at the news cluster denoted by Fig.~\ref{fig:msoverview}G, which \kg{contained} many news articles.
The news cluster was then followed by several Twitter clusters.
After some exploration, \kg{P2} found \kg{that} the news cluster \kg{was} mainly about the first case of Ebola in the US.
The patient, Thomas Duncan, had \kg{been exposed to} as many as 80 people.
The first confirmed case led to \kg{numerous} discussion\kg{s} on Twitter
and created fear.
Because of the heightened attention from the public and media, this topic \docpr{was} divided into four sub-topics:
1) the further report of suspects (Fig.~\ref{fig:msoverview}H);
2) government actions (Fig.~\ref{fig:msoverview}I);
3) treatment of the patient (Fig.~\ref{fig:msoverview}J);
and 4) \docpr{the} search for people who had some form of contact with the patient (Fig.~\ref{fig:msoverview}K).\looseness=-1

\kg{P2} commented that the public in the US \kg{paid minimal} attention to the Ebola epidemic \dc{in} Africa \kg{before Sep. 27, observing the epidemic from the other's perspective.}
This is consistent with the theory of alterity (otherness)~\cite{otherness}.
\kg{When the epidemic} \dc{arrived} \kg{in the US, the perspective changed} and led to \dc{intense} discussions on news media and Twitter.
\kg{P2} further explained \kg{that} the spread of the first case in the US also disclosed another phenomenon.
Since the news media reported the first case wantonly, the severity of the epidemic was overestimated and \docpr{fear was} created among average \dc{people}.
This is \dc{because human perception} is often influenced by the pseudo society built by the media.
Under such \kg{a} situation, the government \kg{must} guide public opinion.

\noindent \textbf{\normalsize \kg{Action} and guidance \kg{of the government}.}
\kg{P2} continued examining new documents to \dc{learn} the actions of the government \dc{regarding} the epidemic.
She found few \dc{discussions} on Twitter on \docpr{the} topic ``\dc{government actions}'' from Oct. 5 to Oct. 11 (Fig.~\ref{fig:ebola}A)\kg{,}
\kg{indicating} that this topic \kg{failed to} attract \kg{public} attention.
On the contrary, \kg{numerous discussions} on Twitter \kg{focused} on the death of \kg{an} Ebola patient (Fig.~\ref{fig:ebola}B).\looseness=-1

\kg{To identify} the reason, \kg{P2} examined these two topics in Fig.~\ref{fig:ebola}(a) and found one representative cluster (Fig.~\ref{fig:ebola}A1, ``EbolaCDC urges hospitals to follow Ebola-related protocols'') and document (Fig.~\ref{fig:ebola}B1, ``Dallas hospital isolating patient being tested for Ebola'').
She then explored similar documents \kg{or} clusters in the adjacent time steps to \dc{determine} \kg{the evolution of} the topics in the stream.
From the links and corresponding documents in Fig.~\ref{fig:ebola}A, \kg{P2} found \kg{that} the government immediately took \kg{action} and prepared for Ebola before Oct. \kg{4}, \kg{as shown by the following news articles}: 1) Sep. 30, ``Health Ministry to distribute 10,000 PPEs on Thursday'' (Fig.~\ref{fig:ebola}A2); 2) Oct. 2, ``Local hospitals prepared in case of Ebola'' (Fig.~\ref{fig:ebola}A3); \kg{and} 3) Oct. 2, ``Ebola `unlikely' but South prepared'' (Fig.~\ref{fig:ebola}A4).
From the links and corresponding clusters in Fig.~\ref{fig:ebola}B, \kg{P2 realized} that the patient's condition \kg{worsened and led to death, as indicated by the following news articles}: 1) Oct. 5, ``Dallas Ebola patient is in critical condition, hospital says'' (Fig.~\ref{fig:ebola}B2); 2) Oct. 5, ``Ebola patient in Dallas takes a turn for worse'' (Fig.~\ref{fig:ebola}B3); \kg{and} 3) Oct. 8, ``Dallas Ebola Patient Dies'' (Fig.~\ref{fig:ebola}B4).
\kg{P2} believed that the government acted promptly.
However, the death of the patient
\kg{led} to uncontrolled public opinion.

\begin{figure*}[t]
	\centering
	\centering
	\includegraphics[width=\linewidth]{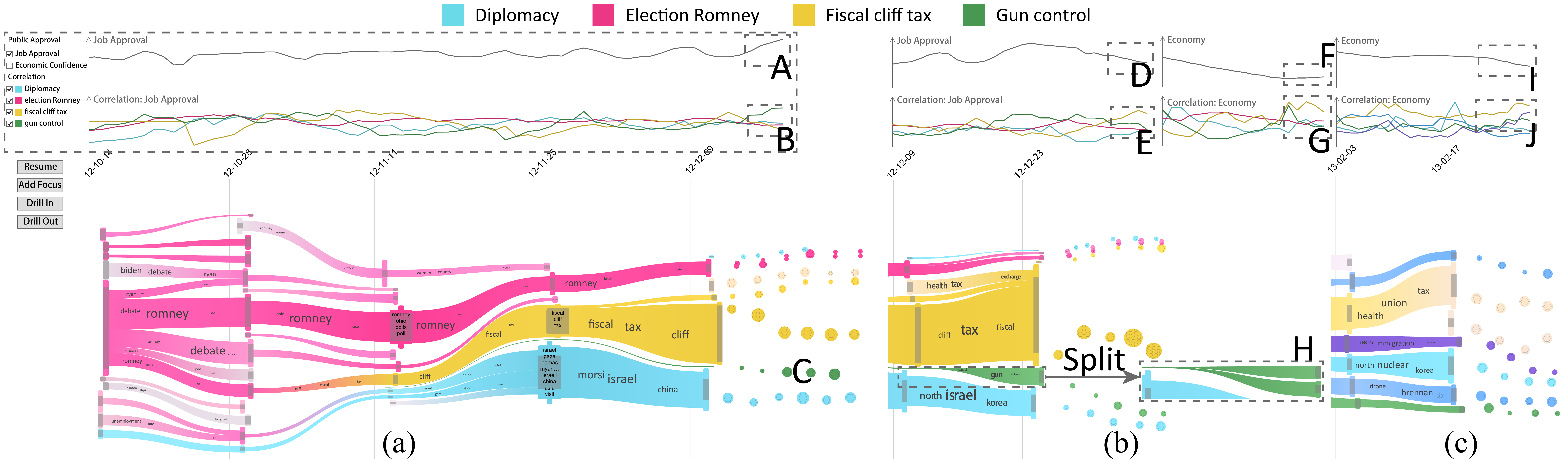}
	\vspace{-4mm}
	\caption{
		Significant changes in public opinion in the Obama dataset.
		(a) Presidential approval \docpr{rating} affected by topic ``Gun Control.''
		(b) A decrease \kg{in} presidential approval and economic confidence caused by the fiscal cliff crisis.
		(c) Another low economic confidence \kg{rating} caused by failed negotiation\kg{s} on \docpr{government spending cuts}.
	}
	\vspace{-5mm}
	\label{fig:obamacase1}
\end{figure*}

\kg{The} documents \kg{that subsequently streamed in were from} the week of Oct. 11.
As shown in Fig.~\ref{fig:ebola}(b), public attention \docpr{to the}  topic ``\dc{government action}'' \kg{decreased} (Fig.~\ref{fig:ebola}E)\kg{, whereas}
discussions on Twitter on topics ``airport screening'' (Fig.~\ref{fig:ebola}C, ``Ebola Screenings Begin at US Airports''), ``travel ban'' (Fig.~\ref{fig:ebola}D, ``RT @CronkiteSays: VIEWER POLL\#N\#Do you support a travel ban from Ebola inflicted countries?''), and ``infected nurse'' (Fig.~\ref{fig:ebola}F, ``Dallas Nurse With Ebola Identified'') \kg{increased}.
The public \kg{paid} \dc{more} attention to negative messages.
The situation \dc{improved} after three weeks (Fig.~\ref{fig:ebola}(c)).
\kg{P2} analyzed the documents and \kg{specified} three reasons for this change.
First, the \kg{change} \dc{from topic} ``airport screening'' \dc{to} another \dc{topic} shifted \dc{public attention} (\kg{Fig.~\ref{fig:ebola}G}).
The new topic \kg{was} \dc{related} to quarantine, \kg{which emerged because} nurse Kaci Hickox \dc{defied} the quarantine \kg{imposed on her} after \kg{returning} from treating Ebola patients in West Africa.
This event caused great disturbance and \docpr{shifted public} attention.
Second, topic ``travel ban'' gradually \kg{disappeared} after \kg{President} Obama decided to cancel the travel ban (Fig.~\ref{fig:ebola}H).
Third, the popularity of ``infected nurse'' gradually decreased as the nurse was cured and \kg{returned} to normal life (Fig.~\ref{fig:ebola}I).
By now, fear caused by the first case of Ebola in the US disappeared and the government finally influenced \dc{public opinion to} \docpr{be more} positive.
\kg{P2} indicated that the government was successful \kg{in} using another topic (``quarantine'', Fig.~\ref{fig:ebola}G) to shift \dc{public attention} \dc{away} from the negative opinion caused by the first \kg{Ebola} case \kg{in the US}.\looseness=-1

\subsection{Obama data}

\begin{table}[b]
	
	\vspace{-3mm}

	\centering
	\scalebox{0.9}{
		\begin{tabular}{|c|c|c|c|c|}
			\hline
			Data & Time span & $N_{num}$ & $h$ & $I_{num}$ \\
			\hline
			\emph{Old} & 10/14/2012-12/8/2012 & 47,963  & 4-10 & 267-376\\
			\hline
			\emph{New} & 12/9/2012-2/21/2015 & 495,151 & 7-11 & 246-471\\
			\hline
		\end{tabular}
	}
	\caption{
		\kg{Statistics} of Obama dataset.
	}
	\label{table:obama}	
\end{table}

The second case study was \kg{a collaboration} with a professor \kg{(}P1\kg{)} \kg{of} media and \docpr{communications}.
In this case study, \kg{P1 studied} the relationship between the media agenda (mass media) and public opinion,
a long-\kg{standing} research topic in \kg{the} field of media and \dc{communication}~\cite{McCombs1972}.

We used a news dataset collected by using \dc{the} keyword ``Obama'' \xiting{(Dataset B)}, which is summarized in Table~\ref{table:obama}.

To analyze the \kg{relationship} between the media agenda and public opinion, \kg{several} \docpr{pieces of} contextual information \dc{were} added (the dashed rectangle in Fig.~\ref{fig:obamacase1}(a)).
The contextual information \dc{consists} of: 1) \kg{Obama's} presidential approval \kg{rating}\kg{,} 2) \kg{an} economic confidence index derived from Gallup public opinion polls~\cite{Gallup}, \kg{and}
3) \dc{a} time\kg{-}varying statistical correlation between \docpr{the} Gallup poll results and the sentiment of media articles.

A word-embedding-based sentiment classification algorithm~\cite{Tang2014} was employed to calculate the sentiment for each article.
The topic \kg{``}sentiment\kg{''} at each time step \kg{refers to} the average sentiments of the documents at that time step.
\kg{A} sentiment time series \kg{was then obtained} for each topic.
Finally we calculated the Pearson correlation coefficient between a Gallup poll result and the temporal sentiment of a topic.\looseness=-1

\noindent \textbf{\normalsize \docpr{The presidential} approval \docpr{rating} affected by \docpr{the} topic ``\kg{gun control}.'' }
In the old data (Fig.~\ref{fig:obamacase1}(a)), \kg{P1} detected four different topics: ``diplomacy'' (blue), ``election'' (red), ``fiscal cliff and taxes'' (yellow), and ``gun control'' (green).
He then started the analysis from Dec. 9, \dc{2012,} \kg{which} was just before the formal re-election of President Obama.
\kg{P1} observed an increase \dc{in the curve for} \docpr{the presidential approval rating} (Fig.~\ref{fig:obamacase1}A).
By comparing the correlation between this index and the sentiment curve of each topic, he found that the highest correlation was with \docpr{the} topic ``gun control'' (Fig.~\ref{fig:obamacase1}B).
This topic received much more attention than \docpr{before the} week of Dec.~9 (Fig.~\ref{fig:obamacase1}C),
which \kg{was} triggered by \dc{a shooting} massacre at \kg{a} Connecticut elementary school on Dec. 14, 2012.
To examine how people responded to this \kg{incident}, P1 split this topic and found two subtopics \kg{(Fig.~\ref{fig:obamacase1}H)}.
One is the president's response and the other is the response of others (\kg{congressional representatives}, NRA, and the public).
\kg{P1} found that the public \dc{called for} tighter gun control (``Gun-control petition to White House breaks record'').
Obama's \dc{actions fit with} public opinion \dc{very well} (``Obama vows to battle gun violence'').
We speculate that this \dc{was} the major cause for the increase \dc{in} his approval rating.\looseness=-1

\begin{figure*}[t]
	\centering
	\includegraphics[width=0.85\linewidth]{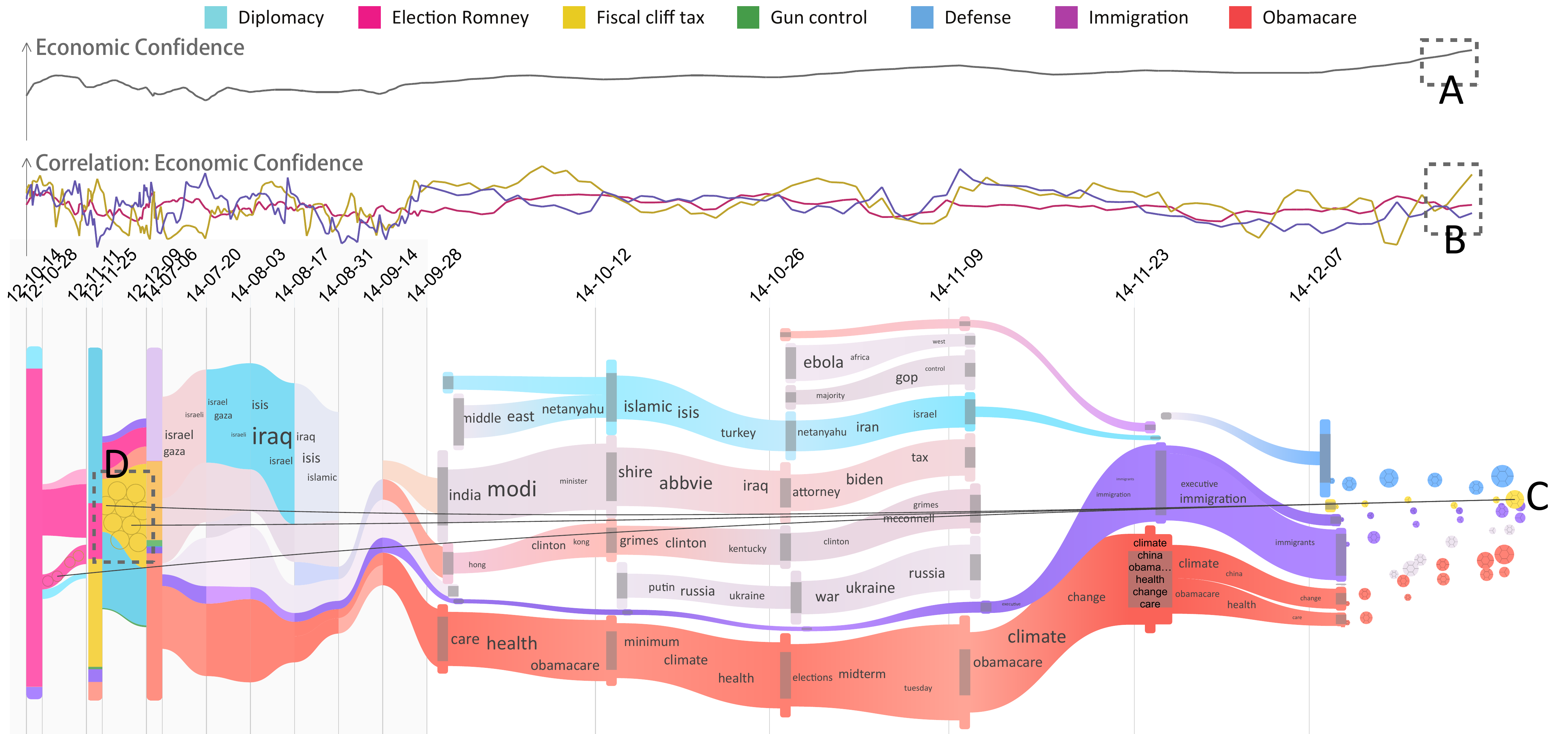}
	\vspace{-2mm}
	\caption{
		Carry-over effect of media agenda: the documents \kg{on} \docpr{the} tax \kg{increase} in 2012 are connected to tax breaks in 2014.\looseness=-1
	}
	\label{fig:obamacase2}
	\vspace{-5mm}
\end{figure*}


\noindent \textbf{\normalsize Public attention \dc{transition} to topic ``\kg{fiscal cliff and taxes}.''}
\kg{P1} observed an immediate decrease \dc{in} \docpr{the \kg{presidential} approval rating} on Dec. 31, 2012 (Fig.~\ref{fig:obamacase1}D).
The correlation between \dc{presidential} approval and \docpr{the} topic ``gun control'' decreased to a smaller value (0.12), \kg{whereas} its correlation with topic ``fiscal cliff and taxes'' increased to \dc{its} highest (0.51, Fig.~\ref{fig:obamacase1}E).
\kg{P1} explained that this topic was \kg{about} the fiscal cliff crisis at the end of 2012.
The government \kg{faced} an act \kg{that would take} effect on Jan. 1, 2013.
\kg{Large} tax \kg{increases} and spending cuts were included in this act.
To postpone this act, the president and the two \kg{political} parties debated \dc{for a long time} and settled \dc{on} a temporary solution on Jan. 1.
They agreed to postpone the spending cuts until Mar. 1.
After reading the news\kg{, P1} found that people \dc{surmised that the president did not truly want the crisis to end}.
As this topic \dc{concerned} \docpr{the} economy, P1 \dc{considered} the economic confidence index.
Unsurprisingly, a local minimum on Dec. 31, 2012 \kg{(Fig.~\ref{fig:obamacase1}F)} \kg{was found}.
The low confidence \kg{level} was possibly caused by raising tax \kg{rates} \kg{because} the correlation between this topic and the economic confidence was \docpr{at its} highest \kg{(Fig.~\ref{fig:obamacase1}G)}.\looseness=-1


As the spending cuts \dc{were} postponed to Mar. 1, \kg{P1} decided to continue tracking this event.
He learned that this act would have a \kg{significant} effect on the economy (``Bernanke: sequester cuts slow economic recovery'').
The spending cuts took effect on Mar. 1.
On \kg{this date}, \kg{P1} observed another local minimum \dc{in} the economic confidence \kg{(Fig.~\ref{fig:obamacase1}I)}.
The correlation between the economic confidence and \docpr{the} topic ``fiscal cliff and taxes" was \dc{at its} highest \kg{(Fig.~\ref{fig:obamacase1}J)},
which \kg{was} in \kg{accordance} with P1's expectation.
He commented that the streaming visualization \kg{was} visually appealing and practically useful  
\kg{for examining} real-time documents.\looseness=-1

\noindent \textbf{\normalsize Carry-over effect of topic ``fiscal cliff and taxes.''}
P1 wanted to follow the subsequent development of this topic.
He found that \docpr{the} topic ``fiscal cliff and taxes" (yellow) appeared again on Dec. 7, 2014 (Fig.~\ref{fig:obamacase2}).
This topic \dc{concerned} tax breaks at the end of 2014. 
At this time, the economic confidence index \dc{experienced} a remarkable increase (Fig.~\ref{fig:obamacase2}A).
Because the correlation between this index and \docpr{the} topic ``fiscal cliff and taxes" was \docpr{at its} highest (0.44, Fig.~\ref{fig:obamacase2}B), P1 speculated that \dc{intense} discussions on tax breaks \dc{concerning} this topic \dc{were} a potential reason.
\kg{P1} was curious \kg{about the} significant influence \kg{of this small topic} on economic confidence.
To this end, \kg{P1} linked the largest document cluster (Fig.~\ref{fig:obamacase2}C) at this time to the previous relevant documents.
\kg{Several} documents appeared during the period of \kg{the} fiscal cliff crisis in 2012 (Fig.~\ref{fig:obamacase2}D).
At that time, the government wanted to raise \kg{taxes because of the} fiscal cliff and this topic was dominant in \kg{the} media (yellow topic in Figs.~\ref{fig:obamacase1}(a) and (b)).
\kg{P1} commented that this \kg{fact} \kg{could} be regarded as \dc{a} carry-over effect \cite{carry-over} in the field of media and \dc{communication}.
\kg{P1} further explained, ``The fiscal cliff crisis left a profound impression on the public and had a great influence \kg{on} the economic confidence at that time.
As a result, this influence can be carried over to the relevant topic later even if it is a smaller one.''

\section{Discussion and Future Work}\label{sec:conclustion}
In this paper, we have presented a novel visual analytics system to help users explore and understand hierarchical topic evolution in high-volume text streams.
Powered by the streaming tree cut model and the corresponding visualization, the system allows users to analyze hierarchical topics at different \dc{granularities,} as well as their evolution patterns over time.
In addition, it \kg{enables} users to interactively customize and refine \dc{the} visualization based on their interests.
A quantitative evaluation and two case studies were conducted to demonstrate the effectiveness and usefulness \kg{of the system} \dc{for} text stream analysis.

Although the system performs well when analyzing the evolution of hierarchical topics, it can still be improved.
First, one component of our system, the evolutionary tree clustering algorithm, is effective in constructing a sequence of topic trees with high fitness and smoothness.
However, relying \kg{solely} on the optimization results is not always effective because the tree clustering algorithm may be imperfect and different users may have different \kg{requirements}.
Studying how to leverage the domain knowledge of a user in our system and allow him/her \kg{to express and define information requirements can help solve the aforementioned problem.}
\kg{This noteworthy topic can be pursued} in the future.
Second, we only \docpr{utilized} the horizontal offset to encode tree depth but \docpr{ignored} the general \docpr{structure} \kg{of a tree}.
However, users may want to examine each tree structure and obtain a complete overview of evolving topic trees \kg{in several cases}.
\kg{We} also plan to enable tree \docpr{exploration in} the next version of the system and allow users to \docpr{explicitly} explore topic hierarchical \docpr{structures.}



\section{Acknowledgements}
This research was supported by National Key Technologies R\&D Program of China (No. 2015BAF23B03), the National Natural Science Foundation of China (No.s 61373070, 61272225, 61572274), and a Microsoft Research Fund (No. FY15-RES-OPP-112). 
}

\small
\bibliographystyle{abbrv}
\bibliography{reference}



\vspace{-10mm}
\begin{IEEEbiography}[{\includegraphics[width=1in,height=1.25in,clip,keepaspectratio]{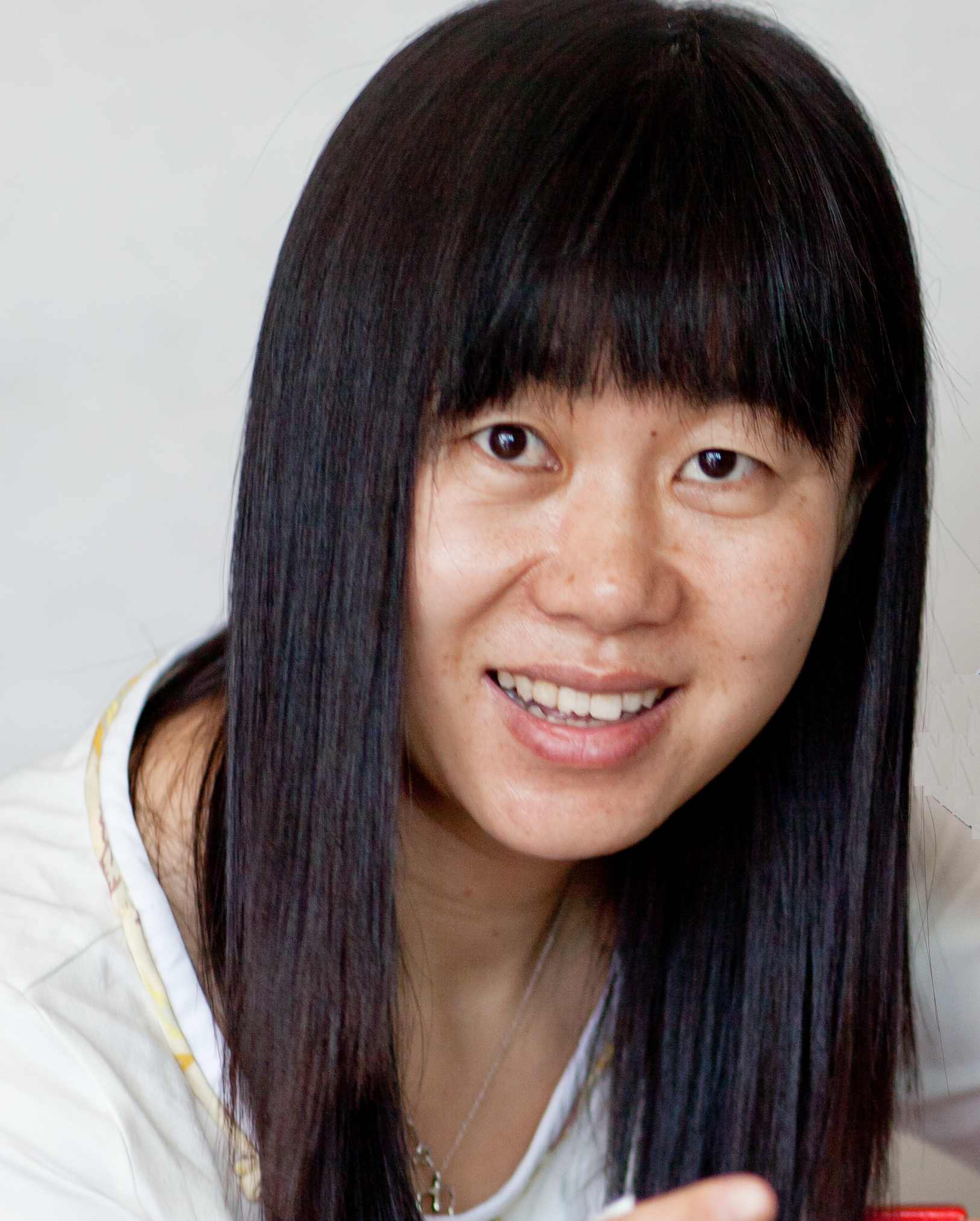}}]{Shixia Liu}
is an associate professor at Tsinghua University. Her research interests include visual text analytics, visual social analytics, and text mining. She worked as a research staff member at IBM China Research Lab and a lead researcher at Microsoft Research Asia.
She received a B.S. and M.S. from Harbin Institute of Technology, a Ph.D. from Tsinghua University. 
She is an associate editor of IEEE TVCG.
\vspace{-8mm}
\end{IEEEbiography}

\vspace{-8mm}
\begin{IEEEbiography}[{\includegraphics[width=1in,height=1.25in]{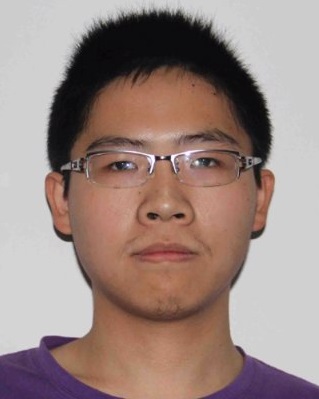}}]{Jialun Yin} is a PhD candidate in the Department of Computer Science and Technology at Tsinghua University, China.
His research interests include visual text analytics and data mining.
He received a BS degree in Computer Science from Tsinghua University.
\vspace{-8mm}
\end{IEEEbiography}

\vspace{-8mm}
\begin{IEEEbiography}[{\includegraphics[width=1in,height=1.25in]{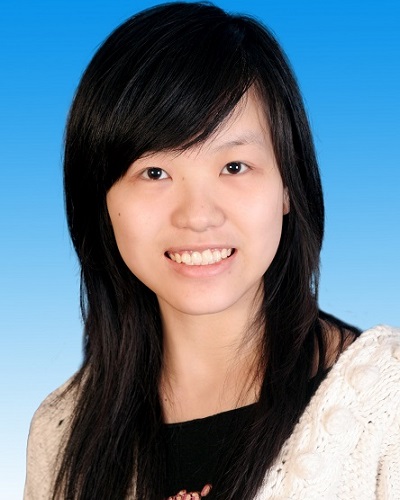}}]{Xiting Wang} is a PhD candidate in the Institute for Advanced Study at Tsinghua University, China.
Her research interests include visual text analytics and text mining.
She received a BS degree in Electronics Engineering from Tsinghua University.
\vspace{-8mm}
\end{IEEEbiography}

\vspace{-8mm}
\begin{IEEEbiography}[{\includegraphics[width=1in,height=1.25in,clip,keepaspectratio]{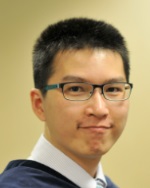}}]{Weiwei Cui}
is a researcher in the Internet Graphics Group at Microsoft Research Asia.
His research interests include visualization and visual analytics, with emphasis on text and graph data.
He received a PhD in computer science from Hong Kong University of Science and Technology.

\vspace{-8mm}
\end{IEEEbiography}

\vspace{-8mm}
\begin{IEEEbiography}[{\includegraphics[width=1in,height=1.25in,clip,keepaspectratio]{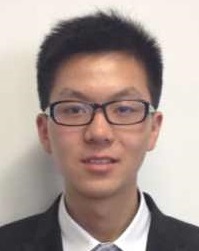}}]{Kelei Cao}
is an undergraduate in the Department of Computer Science and Technology at Tsinghua University, China. His research interests include visual text analytics.

\vspace{-8mm}
\end{IEEEbiography}

\vspace{-8mm}

\begin{IEEEbiography}[{\includegraphics[width=1in,height=1.25in,clip,keepaspectratio]{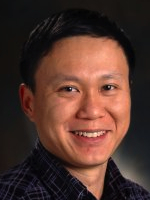}}]{Jian Pei}
is currently the Canada Research Chair (Tier 1) in Big Data Science and a professor at the School of Computing Science and the Department of Statistics and Actuarial Science at Simon Fraser University, Canada. He received his Ph.D. degree at the same school in 2002 under Dr. Jiawei Han's supervision.  His research interests are to develop effective and efficient data analysis techniques for novel data intensive applications.  He has published prolifically and is one of the top cited authors in data mining.  He received a series of prestigious awards.  He is also active in providing consulting service to industry and transferring the research outcome in his group to industry and applications.  He is an editor of several esteemed journals in his areas and a passionate organizer of the premier academic conferences defining the frontiers of the areas. He is an IEEE Fellow.
%
\end{IEEEbiography}

\end{document}